\documentclass[useAMS,usenatbib,usegraphicx,newapa]{mn2e}
\long\def\symbolfootnote[#1]#2{\begingroup%
\def\thefootnote{\fnsymbol{footnote}}\footnote[#1]{#2}\endgroup}

\title[WET observations of EC\,20058]
      {Whole Earth Telescope observations of the hot helium atmosphere 
       pulsating white dwarf EC\,20058$-$5234}
\author[D.~J.~Sullivan et al.]
       {D.~J.~Sullivan,$^{1}$\thanks{}\footnotemark\ 
        T.~S.~Metcalfe,$^{2,3}$
        D.~O'Donoghue,$^4$
        D.~E.~Winget,$^2$
        D.~Kilkenny,$^{4,5}$
        \newauthor
        F.~van Wyk,$^4$
	A.~Kanaan,$^6$
	S.~O.~Kepler,$^7$
        A.~Nitta,$^{2,8,9}$
	S.~D.~Kawaler$^{10}$
        \newauthor
        M.~H.~Montgomery,$^2$
	R.~E.~Nather,$^2$
	M.~S.~O'Brien,$^{10,11}$
        A. Bischoff-Kim,$^2$
        \newauthor
	M.~Wood,$^{12}$
        X.~J.~Jiang,$^{13}$
	E.~M.~Leibowitz,$^{14}$
	P.~Ibbetson,$^{14}$
        S.~Zola,$^{15,16}$
        \newauthor
        J.~Krzesinski,$^{16}$
        G.~Pajdosz,$^{16}$
        G.~Vauclair$^{17}$
        N.~Dolez,$^{17}$
	and M.~Chevreton$^{18}$\\
$^1$School of Chemical \& Physical Sciences, Victoria University of Wellington,
    P.O. Box 600, Wellington, New Zealand.\\
$^2$Department of Astronomy and McDonald Observatory, University of Texas,
    Austin TX 78712, USA.\\
$^3$High Altitude Observatory, National Center for Atmospheric Research,
    P.O. Box 3000, Boulder, CO 80307, USA.\\
$^4$South African Astronomical Observatory, P.O. Box 9, Observatory 7935, 
    South Africa.\\
$^5$Department of Physics, University of the Western Cape, Private Bag X17,
    Belville 7535, South Africa.\\
$^6$Departamento de F\'{i}sica, UFSC, CP 476, 88040-900 Florian\'{o}plois SC,
    Brazil.\\
$^7$Instituto de F\'{i}sica da UFRGS, 91501-900 Porto Alegre RS, Brazil.\\
$^8$Visiting astronomer, Cerro Tololo Inter-American Observatory, Chile.\\
$^9$Gemini Observatory, 670 N A'ohoku Pl., Hilo, HI 96720, USA\\
$^{10}$Department of Physics \& Astronomy, Iowa State University, Ames,
       IA 50011 USA.\\
$^{11}$Department of Astronomy, Yale University, P.O. Box 208101, New Haven,
       CT 06511, USA.\\
$^{12}$Department of Physics and Space Sciences and SARA Observatory,
       Florida Institute of Technology, Melbourne, FL 32901-6975, USA.\\
$^{13}$Beijing Astronomical Observatory, Chinese Academy of Sciences,
       20 Datun Road, Chaoyang, Beijing 100012, China.\\
$^{14}$Department of Physics and Astronomy and Wise Observatory, Tel Aviv 
       University, Tel Aviv 69978, Israel.\\
$^{15}$Astronomical Observatory, Jagiellonian University, Ul. Orla 171,
       30-244 Cracow, Poland.\\
$^{16}$Mount Suhora Observatory, Pedagogical University, Ul. Podchoraazych 2,
       30-024 Cracow, Poland.\\
$^{17}$Observatoire Midi-Pyr\'{e}n\'{e}es, Universit\'{e} Paul Sabatier,
       CNRS/UMR5572, 14 Avenue Edouard Belin, 31400 Toulouse, France.\\
$^{18}$Observatoire de Paris-Meudon, LESIA, 92195 Meudon, France.}

\date{Accepted 2008 February 5. Received 2008 February 5; In original form 2007 November 7}

\begin{document}

\label{firstpage}

\maketitle

\begin{abstract}
  We present the analysis of a total of 177\,h of high-quality optical
  time-series photometry of the helium atmosphere pulsating white
  dwarf (DBV) EC\,20058$-$5234.  The bulk of the observations (135\,h)
  were obtained during a WET campaign ({\sc xcov}15) in July 1997 that
  featured coordinated observing from 4 southern observatory sites
  over an 8-day period.  The remaining data (42\,h) were obtained in
  June 2004 at Mt John Observatory in NZ over a one-week observing
  period.  This work significantly extends the discovery observations
  of this low-amplitude (few percent) pulsator by increasing the
  number of detected frequencies from 8 to 18, and employs a
  simulation procedure to confirm the reality of these frequencies to
  a high level of significance (1 in 1000).  The nature of the
  observed pulsation spectrum precludes identification of unique
  pulsation mode properties using any clearly discernable trends.
  However, we have used a global modelling procedure employing genetic
  algorithm techniques to identify the $n$, $\ell$ values of 8
  pulsation modes, and thereby obtain asteroseismic measurements of
  several model parameters, including the stellar mass (0.55
  M$_{\sun}$) and T$_{\rm eff}$ ($\sim$ 28\,200\,K).  These values are
  consistent with those derived from published spectral fitting:
  T$_{\rm eff} \sim$ 28\,400\,K and $\log g \sim$ 7.86.  We also
  present persuasive evidence from apparent rotational mode splitting
  for two of the modes that indicates this compact object is a
  relatively rapid rotator with a period of 2\,h.  In direct analogy
  with the corresponding properties of the hydrogen (DAV) atmosphere
  pulsators, the stable low-amplitude pulsation behaviour of EC\,20058
  is entirely consistent with its inferred effective temperature,
  which indicates it is close to the blue edge of the DBV instability
  strip.  Arguably, our most significant result from this work is the
  clear demonstration that EC\,20058 is a very stable pulsator with
  several dominant pulsation modes that can be monitored for their
  long term stability.


\end{abstract}

\begin{keywords}
  stars: individual: EC\,20058$-$5238 -- white dwarfs --
  stars: oscillations -- stars: interiors -- techniques: photometric
\end{keywords}

\section{Introduction}

Nearly 99\% of all stars are predicted to end their lives as slowly
cooling white dwarfs; their study, among other things, provides an
important perspective on the active lives of all the progenitor
objects.  In particular, if we can determine their internal chemical
compositions, we have access to key data concerning the products of
the nuclear reactions that power stars in the previous evolutionary
stages.  In common with other astronomical objects, we are largely
limited to studying the atmospheres of white dwarfs as this is where
the photons we detect are created.  Thus, detailed spectroscopic
observations allow us to measure effective temperatures, surface
gravity values (and therefore stellar masses), and atmospheric
chemical compositions.  From these studies over many years we know
that the great majority of white dwarfs divide into two classes
\citep[e.g.][]{mcc99}.  The DA class have hydrogen atmospheres, and
the DB class have \emph{pure} helium atmospheres (no detectable sign
of hydrogen and other elements).  The DAs account for about 86\% of
all white dwarfs, while the DBs dominate the rest.
\symbolfootnote[0]{*\ E-mail: denis.sullivan@vuw.ac.nz}
\symbolfootnote[0]{\dag\ Visiting astronomer, Mt John University Observatory, 
operated by the Department of Physics \& Astronomy, University of Canterbury.}

In both of these classes there are objects that provide further
information that is subtly encoded in stellar flux variations. These
are the pulsators.  The cooler hydrogen atmosphere pulsators (DAV
class) were discovered serendipitously in the late 1960s \citep{lan68},
while their hotter helium atmosphere cousins were detected following
theoretical predictions and a targeted search \citep{win81,win82}.

The use of pulsating stars to infer some of their intrinsic properties
is called stellar seismology or simply \emph{asteroseismology}.
Asteroseismology uses the detected pulsation modes of a star to
constrain computer models and thereby measure stellar properties,
including otherwise hidden interior physical quantities.  This
endeavour has been a very productive exercise \citep[e.g.][]{win98} for
white dwarfs, in part due to the relative simplicity of the white
dwarf structure, but also due to the potentially rich pulsation mode
spectrum that results from the nonradial spheroidal (g-mode) pulsation
mechanism.  All observed white dwarf pulsations are attributed to
buoyancy-driven g-modes, as the observed periods (70\,s -- 1500\,s)
are inconsistent with the predicted periods of pressure (acoustic)
waves in the extremely dense white dwarf material: such pulsations
would have periods several orders of magnitude smaller, and have never
been observed inspite of a number of searches \citep{rob84,kaw94,sil07}.


The more identified pulsation modes, the more successful is the white
dwarf asteroseismology: in effect, each detected pulsation mode adds
another constraint to the modelling process.

Given that the DA objects form a substantial majority of all white
dwarfs, it is natural that the DAVs dominate the known objects for
these compact pulsators.  In fact, following a recent large increase
in the detected number of faint white dwarfs \citep{kle04}, and
follow-up programmes to detect more pulsators in these new objects
\citep{muk04a,mul05,kep05b,cas07}, the number of hydrogen atmosphere
pulsators has mushroomed to more than 140.

The helium atmosphere DBV pulsators are far fewer: in 2007 there are
17 known objects in this class \citep{bea99,han01,nit05,nit07}.
The southern object EC\,20058$-$5234 (QU Tel, henceforth simply
EC\,20058) was the 8th DBV to be discovered.  It has a B magnitude of
approximatley 15 and was discovered by the Edinburgh-Cape (EC) faint
blue object survey \citep{sto97}.  \citet{koe95} first reported a
study of its properties: they spectroscopically established its DB
classification and also demonstrated its variability using the
techniques of time-series photometry.  Their 20 hours of photometry
over a four month period in 1994 revealed a total of 8 pulsation
frequencies and showed that the object was a low amplitude DBV
variable that appeared to be quite stable; this is in complete
contrast (on both counts) to the class prototype, GD\,358.  In fact
GD\,358 is a large amplitude variable which also exhibits considerable
changes in its observed pulsation spectrum over various timescales
ranging from days to years \citep{kep03}.

\begin{table*}
  \caption{
    Journal of observations of time-series photometry of
    EC\,20058$-$5234 obtained during the Whole Earth Telescope (WET) extended
    coverage campaign ({\sc xcov}15) in July 1997.  Column 2 provides the
    observatory and telescope for each run, which are:
    CTIO = Cerro Tololo Interamerican Observatory, La Serena, Chile 
    (observer: A.\ Nitta); MJUO = Mount John University Observatory,
    Lake Tekapo, New Zealand (observer: D.\ J.\ Sullivan);
    OPD = Observatorio Pico dos 
    Dias, Itajub\'{a}, Brazil (observer A.\ Kanaan);
    and SAAO = South African Astronomical Observatory,
    Sutherland, South Africa (observers: D.\ Kilkenny, F.\ van Wyk).
    Columns 3 and 4 provide the UT starting date and time for each run, 
    column 5 gives the total number of 10 s integrations in each run,
    column 6 gives the run length in hours,
    while the last column is the barycentric Julian start date 
    (BJD$^{-}$ = BJD $-$ 2450000, see text)
    for each run corresponding to the UT start times. Note that the
    two OPD runs marked with an asterisk in column 1 (ra411 \& ra413) were
    not used in the final analysis due to an unexplained timing error (see
    text)}.

\label{tab:x15}
\vspace{5mm}
\begin{tabular}{lrlrrll}
\hline
\multicolumn{1}{c}{Run name}    &
\multicolumn{1}{c}{Telescope}    &
\multicolumn{1}{c}{Date}    &
\multicolumn{1}{c}{UT}    &
\multicolumn{1}{c}{N}    &
\multicolumn{1}{c}{$\Delta$T}    &
\multicolumn{1}{c}{BJD$^{-}$}    \\
\multicolumn{1}{c}{}    &
\multicolumn{1}{c}{}    &
\multicolumn{1}{c}{(1997)} &
\multicolumn{1}{c}{start}    &
\multicolumn{1}{c}{}    &
\multicolumn{1}{c}{[h]}    &
\multicolumn{1}{c}{start}    \\
\hline
dmk057   & SAAO 1.0m & July 2  & 20:05:27 & 2745 & 7.63 & 632.3427230 \\
dmk060   & SAAO 1.0m & July 3  & 19:50:34 & 2762 & 7.67 & 633.3324031 \\
ra404    &  OPD 1.6m & July 4  &  2:28:00 &  448 & 1.24 & 633.6084026 \\
jl0497q1 & MJUO 1.0m & July 4  &  7:30:00 &  402 & 1.12 & 633.8181279 \\
jl0497q2 & MJUO 1.0m & July 4  &  8:57:30 & 2687 & 7.46 & 633.8788927 \\
jl0497q3 & MJUO 1.0m & July 4  & 16:56:00 &  591 & 1.64 & 634.2111891 \\
dmk061   & SAAO 1.0m & July 4  & 20:16:08 &  384 & 1.07 & 634.3501725 \\
dmk062   & SAAO 1.0m & July 4  & 22:07:39 & 1202 & 3.34 & 634.4276157 \\
dmk063   & SAAO 1.0m & July 5  &  1:42:07 &  436 & 1.21 & 634.5765529 \\
dmk064   & SAAO 1.0m & July 5  &  3:00:50 &  352 & 0.98 & 634.6312180 \\
ra407    & SAAO 1.0m & July 5  &  5:33:10 &  996 & 2.77 & 634.7370064 \\
jl0597q1 & MJUO 1.0m & July 5  &  7:26:30 & 1394 & 3.87 & 634.8157112 \\
jl0597q2 & MJUO 1.0m & July 5  & 11:39:00 & 1112 & 3.09 & 634.9910607 \\
jl0597q3 & MJUO 1.0m & July 5  & 14:57:00 &  684 & 1.90 & 634.1285624 \\
jl0597q4 & MJUO 1.0m & July 5  & 17:04:20 &  402 & 1.12 & 634.2169895 \\
jl0597q5 & MJUO 1.0m & July 5  & 18:18:00 &  160 & 0.44 & 634.2681475 \\
dmk066   & SAAO 1.0m & July 5  & 19:57:00 & 2834 & 7.87 & 635.3368984 \\
an-0067  & CTIO 1.5m & July 6  &  2:43:00 & 2579 & 7.16 & 635.6188463 \\
jl0697q1 & MJUO 1.0m & July 6  &  7:31:50 & 3148 & 8.74 & 635.8194274 \\
jl0697q2 & MJUO 1.0m & July 6  & 16:50:40 &  656 & 1.82 & 636.2075105 \\
dmk069   & SAAO 1.0m & July 6  & 19:28:06 & 2994 & 8.32 & 636.3168405 \\
ra410    &  OPD 1.6m & July 7  &  0:28:40 & 1913 & 5.31 & 636.5255696 \\
jl0797q1 & MJUO 1.0m & July 7  &  7:16:30 & 2211 & 6.14 & 636.8087902 \\
jl0797q2 & MJUO 1.0m & July 7  & 13:46:00 &  782 & 2.17 & 636.0792790 \\
jl0797q3 & MJUO 1.0m & July 7  & 16:08:00 &  809 & 2.25 & 636.1778911 \\
dmk071   & SAAO 1.0m & July 7  & 19:25:06 & 2956 & 8.21 & 637.3147674 \\
ra411 (*)&  OPD 1.6m & July 8  &  4:24:20 & 1181 & 3.28 & 637.6892385 \\
jl0897q1 & MJUO 1.0m & July 8  & 11:03:50 &  291 & 0.81 & 637.9666716 \\
fvw094   & SAAO 1.0m & July 8  & 20:44:00 &  312 & 0.87 & 638.3695685 \\
fvw095   & SAAO 1.0m & July 8  & 23:31:50 & 1567 & 4.35 & 638.4861204 \\
ra413 (*)&  OPD 1.6m & July 9  &  1:35:20 & 1389 & 3.86 & 638.5718849 \\
fvw097   & SAAO 1.0m & July 9  & 19:09:00 & 3033 & 8.43 & 639.3036033 \\
fvw099   & SAAO 1.0m & July 10 & 19:09:10 & 3045 & 8.46 & 640.3037252 \\
\hline
\end{tabular}
\end{table*}

The Fourier analysis of the discovery data set suggested the presence
of low level frequencies below the chosen significance level, so one
of us (DOD) proposed a Whole Earth Telescope (WET) campaign
\citep{nat90} to probe for coherent periodicities at lower amplitudes
than those detected by \citeauthor{koe95}  Consequently, a WET run was
scheduled for dark time in July 1997, corresponding to the optimal
observing season for the target.  Following the WET run, regular
single-site monitoring has been carried out at primarly Mt John
Observatory (NZ).

\section{Observations}

\subsection{WET 1997 photometry}

Observing time covering a period of 9 days on 4 southern telescopes
was obtained in July 1997 in order to monitor EC\,20058.  Time-series
aperture photometry was carried out using photometers equipped with
blue sensitive photomultiplier tubes and no filters in the light beam
at all 4 observing sites.  Consequently, the resulting `white light'
passbands employed in the observations had effective wavelengths
similar to that of Johnson B, but with a significantly wider passband.

All photometers used 10\,s integration times and operated with
essentially 100\% duty cycles, given the nature of the photon counting
systems employed.  The 1.0-m telescope at the South African
Astronomical Observatory (SAAO) used a single channel photometer in
combination with autoguiding; both the 1.0-m telescope at Mt John
University Observatory (MJUO) in New Zealand and the 1.6-m Itajuba
telescope in Brazil were equipped with two-channel photometers; and a
three-channel photometer \citep*{kle96} was used in combination with
the CTIO 1.5-m telescope in Chile.  A journal of all the observations
is provided in Table~\ref{tab:x15}

\begin{figure*}
\includegraphics[width=10.0cm, angle=-90]{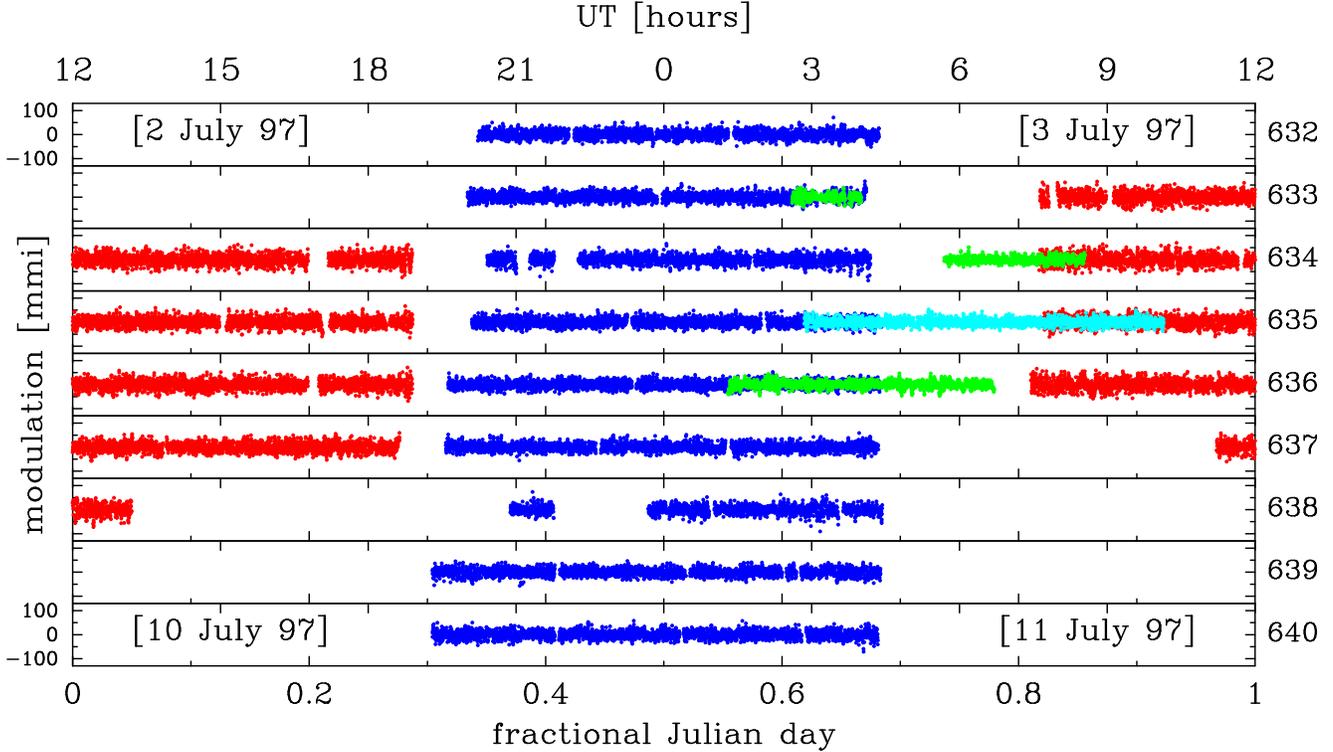}
\caption{The reduced {\sc xcov}15 time-series photometry showing the
  data quality and the extended coverage provided by the multi-site
  campaign.  Each panel represents a whole Julian day with the actual
  day determined by adding 2450000 to the number on the right of each
  panel.  The vertical axes are in units of millimodulation intensity
  [mmi] whereby 10 mmi corresponds to a 1\% flux variation from the
  local mean value.  The (blue) values in each panel centred around
  0\,h UT represent SAAO data, the (red) values centred around 12\,h
  UT represent MJUO data, while the CTIO data (light blue) is centred
  around 6\,h UT in the 4th panel from the top and the OPD data
  (green) segments appear in panels 2,3 and 5 in the vicinity of 6 h
  UT.  Note that the two OPD runs ra411 and ra413 have not been included 
  in the plot, as they were not included in the final analysis (see
  discussion in the text).  Also, see the online journal article for a
  colour version of this figure.}
\label{fig:ts}
\end{figure*}

Good weather at both SAAO and Mt John, combined with poor conditions
at CTIO, meant that the bulk of the observations were obtained at just
two observatories.  However, inspite of the participation of only 4
telescope sites overall, the WET campaign achieved an observing duty
cycle of 64\% for the whole 8.3 day observing period, and a duty cycle
of 89\% for the central 4.4 days when all telescopes contributed.  The
WET campaign complete coverage can be gleaned from Fig.~\ref{fig:ts},
where the reduced photometry has been graphed in successive 24 hour
segments.

\subsection{Mt John 2004 photometry}

Since the 1997 WET campaign, EC\,20058 has been regularly monitored
from Mt John during its extended southern observing season by one of
us (DJS), and some preliminary reports on this work have already
appeared \citep{sul00b,sul03}.  Quality time-series photometry
obtained in June 2004 provided confirmation of a pulsation mode that
is close to marginal in the WET data set, so it is germane to include
some analysis of these data here.

The Mt John 1.0-m telescope in combination with a three-channel
photometer \citep*{kle96} using unfiltered `white light' (and hence a
passband similar to that mentioned previously) was used to acquire
time-series photometry of EC\,20058 in June 2004.  This photometer
includes the functionality of the one used at CTIO in the 1997 WET run,
which enables the observer to monitor 3 regions of the sky using
nominally identical miniature blue sensitive (Hamamatsu R647-04)
photomultiplier tubes.  But, it also incorporates a very useful
improvement \citep[e.g.][]{sul00a}.  In addition to continously
monitoring the target star, a comparison star and sky background, the
comparison star can also function as a continuous guide star.  This is
achieved by using a dichroic filter in the comparison star's optical
train so that the red component of its spectrum is transmitted to a
small CCD and the remaining blue component is reflected to the
photomultiplier tube.  Continuous remote guiding or autoguiding is
then possible using only the capabilities of the photometer.

\begin{table}
  \caption{Journal of observations of time-series photometry of
    EC\,20058$-$5234 obtained at Mt John Observatory in June 2004.
    Columns 3 and 6 give the UT and (modified) barycentric Julian date
    (BJD$^{-}$) start times for each run and columns 4 and 5 provide
    the number of useful 10 s integrations and length of run in hours,
    respectively.}
  \label{tab:mj}
  \vspace{5mm}
\begin{tabular}{llrrrll}
\hline
\multicolumn{1}{c}{Run}    &
\multicolumn{1}{c}{Date}    &
\multicolumn{1}{c}{UT}    &
\multicolumn{1}{c}{N}    &
\multicolumn{1}{c}{$\Delta$T}    &
\multicolumn{1}{c}{BJD$^{-}$}    \\
\multicolumn{1}{c}{name}    &
\multicolumn{1}{c}{(2004)} &
\multicolumn{1}{c}{start}    &
\multicolumn{1}{c}{}    &
\multicolumn{1}{c}{[h]}    &
\multicolumn{1}{c}{start}    \\
\hline
ju0904q1 & June 9  & 13:18:30 &  827 & 2.97 & 3166.0594447 \\
ju1004q2 & June 10 & 12:15:30 & 2376 & 6.63 & 3167.0157386 \\
ju1104q2 & June 11 &  9:43:10 &  399 & 1.15 & 3167.9099916 \\
ju1104q3 & June 11 & 11:06:10 & 2718 & 7.57 & 3167.9676330 \\
ju1204q2 & June 12 &  9:47:20 & 2010 & 5.61 & 3168.9129289 \\
ju1204q3 & June 12 & 15:29:10 & 1228 & 3.43 & 3169.1503234 \\
ju1404q2 & June 14 & 12:19:20 & 1929 & 6.13 & 3171.0185725 \\
ju1604q2 & June 16 &  9:45:20 & 3067 & 8.54 & 3172.9117024 \\
\hline
\end{tabular}
\end{table}

All of the 2004 Mt John photometry was obtained using the three
channel photometer operated in autoguiding mode.  A good run of clear
weather in June led to a 24\% duty cycle out of a total observing period
of 7.2 days. The results of this work are summarised in Table~\ref{tab:mj}.

\section{Data Reduction} 

We wish to identify periodicities in the light curve that result from
stellar pulsation.  The first stage in this process is to convert the
time series data to a form that emphasises the intrinsic stellar
intensity changes: this requires the removal of a number of artifacts
that result from the observing process.  These reductions steps are
relatively straight forward \citep[e.g.][]{nat90, kep93}, but in the
interests of completeness, and also because we have employed a novel
smoothing procedure, we will briefly summarise our steps here.

The EC\,20058 field of view has two \emph{faint} companions separated
from the target by 2 and 4 arcseconds, respectively.  For the aperture
photometry employed in the work presented here, it was impractical to
attempt to separate the target stars from its two companions
(especially in indifferent seeing conditions), so all observers used
aperture sizes large enough to include all 3 stars and prevent any
errors introduced by imperfect tracking.  It is relevant to note at
this point that the light curve pulsation amplitudes presented later
have not been corrected for this companion star contamination, which
has been estimated by \citet{koe95} to require multiplication by
a factor $\sim$1.4.

For each light curve, the sky background (approaching 50\% in some
cases) for each integration was estimated and then subtracted point by
point from the raw light curves.  For both the single channel and two
channel photometers, sky measurements were made by moving the
telescope to a blank section of sky at the beginning and end of each
run, as well as several times per hour.  A sky background time series
curve was constructed for each light curve by using either linear or
cubic spline interpolation between the measured sky points and then
subtracting this computed curve from the raw light curve.

For the three channel photometers (CTIO in the WET run and all of the
2004 Mt John photometry), the relative channel sensitivities were
calibrated by measuring sky values for all 3 channels at the beginning
and end of each run.  Appropriately adjusted sky values were then
subtracted point by point from the other two raw light curves.

The effects of airmass changes were accounted for by fitting low-order
polynomials to the sky-corrected data, and then determining fractional
deviations of the data points from the fitted values.  Although the
impact of airmass changes can be readily modelled using a
plane-parallel atmosphere approximation, it is just as effective to
use the polynomial fitting method.

Accounting for other light-curve artifacts such as the effect of cloud
is more problematical, but can be achieved for the two and three
channel photometer data by normalising the sky-corrected target data
to the sky-corrected comparison data for those regions affected by
cloud.  The effect of thin cirrus cloud on a moonless night up to a
reduction amount of about 30\% can normally be corrected for.  This
applies in particular to data acquired with the three channel
photometers (such as that obtained at Mt John in 2004) since (a) all
three channels employ identical blue-sensitive photomultiplier tubes
(Hamamatsu R647-04), and (b) the dichroic optical element in the
comparison channel makes its passband effectively similar to that of
the target channel (when viewing the hot white dwarfs), irrespective
of the spectral type of the comparison star.  However, the overriding
criterion employed here is that if a data segment appears irreparably
cloud-contaminated then it is not included in the combined data set.
The data sets listed in Tables \ref{tab:x15} and \ref{tab:mj} adhere
strictly to this criterion.

Any residual `long period' variations in each reduced light curve
segment (now expressed as fractional deviations from the local mean)
were removed using techniques incorporated in the program
\emph{ts3fix}\footnote{http://whitedwarf.org/ts3fix}, developed by one
of us (DJS).  This program allows the user to fit cubic splines to the
data by marking a displayed plot with points at times and flux values
selected by the user.  The flux fitting values can be determined
either by using a local flux average value, or by direct visual
selection. Cubic spline fits to these points are then used to smooth
the light curve over the entire data segment time interval.

Undoubtedly, there is an increased component of subjectivity in this
procedure when compared with an alternative more objective technique
(such as high pass filtering in the frequency domain); the latter can
be readily replicated by others.  One might liken it to `chi-by eye'
visual fitting of an elementary function to a data set compared to use
of a more objective fitting procedure, such as least squares.  Still,
the power of an experienced observer viewing data presented in
suitable graphical form and making informed judgements about the
quality should not be underestimated; experience has shown that these
smoothing procedures are very useful in removing light-curve artifacts
that are clearly not related to the WD pulsations of interest.

However, the procedure is obviously open to `abuse' and consequential
loss (or gain) of signal.  An important option, which acts as a
safeguard, allows the user to calculate Fourier transforms of both the
input and modified data and view a comparison plot in order to monitor
directly in the frequency domain the changes that are made in the time
domain.  The procedures can be easily modified or restarted if one is
suspicious of the changes made to the light curve.  This program has
proved to be very effective in eliminating in the time domain obvious
artifacts that are not related to white dwarf pulsation, such as
inadequate twilight sky correction, and uncorrected transparency
variations due to both airmass changes and cloud.

The resulting corrected light curves are then, of course, blind to
periodic variations larger than a certain period limit, but it is
preferable to remove extraneous power in the time domain rather than
simply rely on signal orthogonality in the Fourier domain to
differentiate between the signals of interest and other artifacts.
Also, given that the white dwarf pulsations of interest are in the
range of about 100\,s to 1000\,s, then if one only makes careful use
of the program's capabilities on longer data segments (at least an
hour or more), then there is a negligible chance of removing any real
signal.  Besides, previous work \citep{bre95,bre96} has demonstrated
that use of the continuous monitoring time-series photometry reported
here is inferior to the more traditional `three star' photometry when
studying longer period ($\sim$ hour or more) variables, such as the
$\delta$ Scuti stars.

Finally, the observation \emph{start times} for each observatory
significant data set were converted to the uniform barycentric Julian
day (BJD) timescale, which corresponds to international time (TAI) in
units of days transformed to the centre of mass of the solar system
\citep{aud01,sta98}.

It is relevant to note here that individual integration times within
each listed data segment were not transformed to the BJD timescale.
This results in a maximum timing discrepancy of about a second (1.2\,s
for the 7.57\,h 2004 Mt John June 11 data), so this omission will have
a negligible impact on the Fourier analysis of the 10\,s integration
time-series data.  However, these timescale changes are taken into
account in software we have developed to search for long term period
changes.

\begin{figure*}
\includegraphics[width=11.5cm, angle=-90]{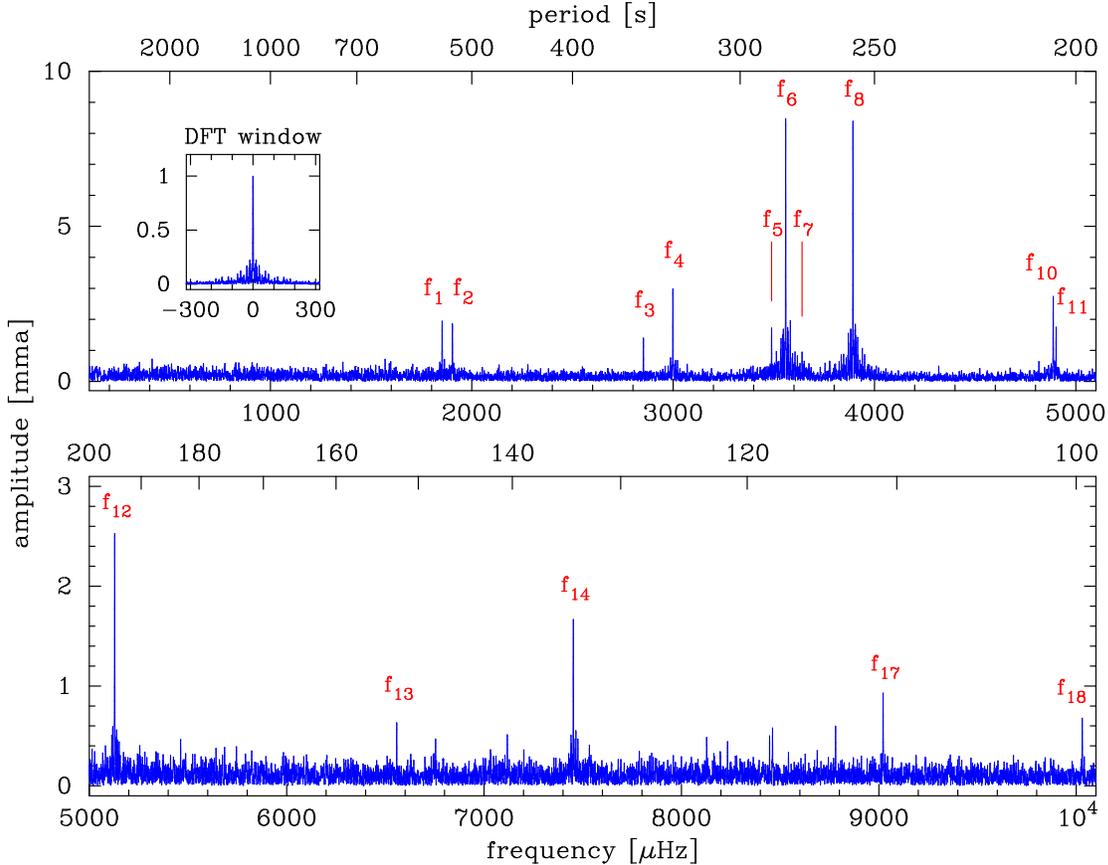}
\caption{An amplitude periodogram (DFT) of the {\sc xcov}15 data set
  together with the DFT window (plot insert) using the same horizontal
  scale as the main plots.  The vertical axes employ the (linear) unit
  of millimodulation amplitude [mma] in which 10 mma corresponds to a
  1\% amplitude modulation of the light flux.  This plot makes clear
  that the EC20058 is a multi-periodic low amplitude pulsator as the
  two dominant modes have amplitides less than 10 mma.  Readily
  identifiable periodicities (15 in all) are labelled and there is a
  suggestion of more real power above 6500$\mu$Hz other than the 4
  labelled peaks f$_{13}$, f$_{14}$, f$_{17}$ and f$_{18}$.  Note that
  f$_7$ is readily identified either by comparing the structure in the
  DFT at frequencies just above the main f$_6$ peak with the window
  function, or (better still) by inspecting a suitably prewhitened DFT
  (Fig. 3).}
\label{fig:dft}
\end{figure*}

\section{Fourier analysis}

The standard way to identify the frequency structure in a reduced
light curve when the phase information is not required is to calculate
a power spectrum.  This is presented for the WET data in
Fig.~\ref{fig:dft} in the form of an amplitude `periodogram' covering
the frequency range from 100 to 10,100 $\mu$Hz, and in which the
vertical axes employ the scale millimodulation amplitude [mma].  We
have employed an amplitude scale rather than a power scale (which is
now common practice) to represent signal `power', as this facilitates a
more direct intuitive connection with actual sinusoidal variations in
the light curve and comparisons with least squares fitting of sinusoidal
functions.  Following the work of \citet{sca82}, a number of authors
use the term `amplitude periodogram' or simply `periodogram' for these
plots.  We will henceforth simply use the acronym DFT (for discrete
Fourier transform) to indicate this procedure.

The WET data cover a period of 8.3 days, which corresponds to a
frequency resolution of $\sim$ 1.4 $\mu$Hz.  The insert panel in
Fig.~\ref{fig:dft} illustrates this resolution by plotting the DFT
`window' for the transform using the same horizontal scale width as
the main plot.  This window corresponds to the DFT of a synthetic
noise-free sinusoid calculated at the same times as all the light
curve data: it directly illustrates the frequency resolution
\emph{and} the impact of the spurious side lobes, or `aliases', that
result from turning the signal on and off during the observing period.
The central plot value has been shifted to zero frequency for clarity.
The near continuous WET data has minimal alias interference, which is,
of course, the primary reason for the observation strategy pursued by
the WET collaboration.  Note that in the wider Fourier literature, the
additional peaks in a DFT deriving from a particular observing window
are termed `spectral leakage', while the term `alias' is reserved for
effects near the Nyquist frequency resulting from undersampled data.
Since we are considering frequencies in our well-sampled data much
smaller than the Nyquist frequency (50,000 $\mu$Hz), following other
authors, we will use the term `alias' for this phenomenon, as it more
accurately describes the additional forest of sometimes confusing
peaks that occur, particularly for extended data sets with regular
daily gaps (eg, see the right hand panels in Figs.~\ref{fig:f6+f8} and
\ref{fig:f10+f11}, corresponding to the MJ04 data).

\begin{figure*}
\includegraphics[width=11.5cm, angle=-90]{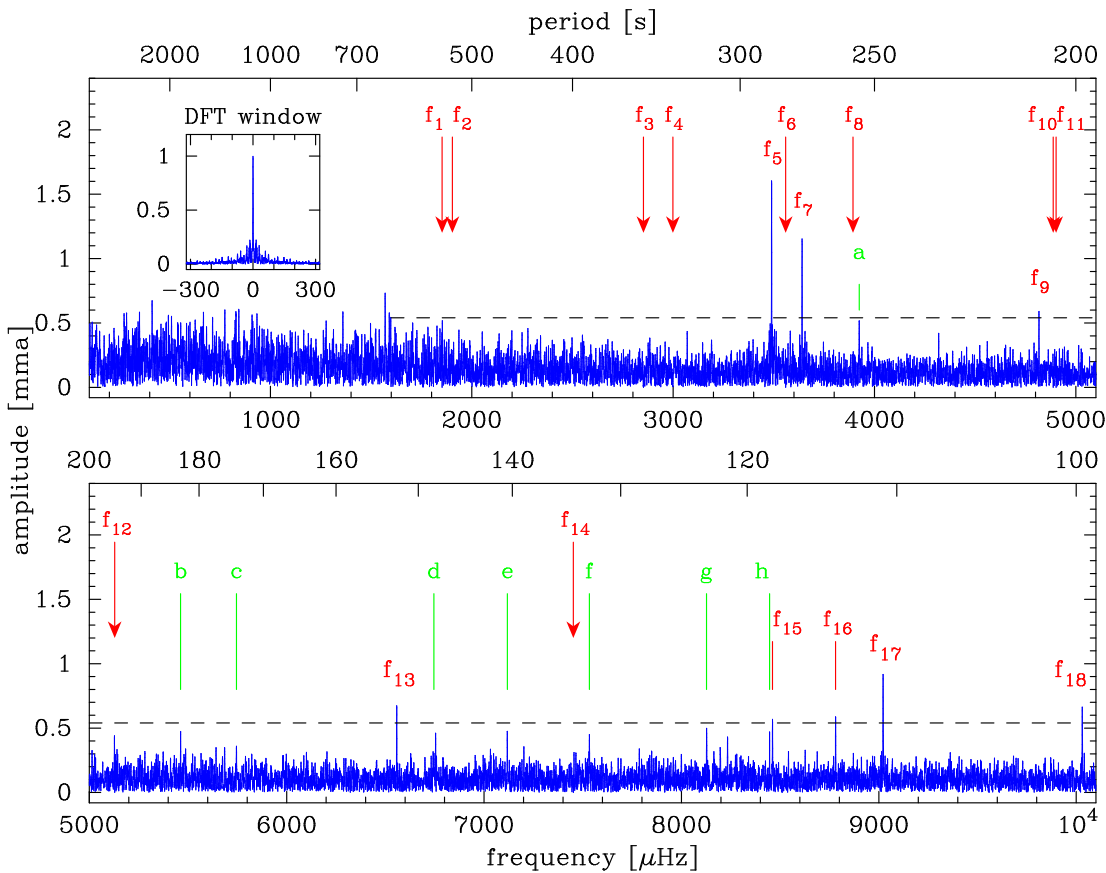}
\caption{The DFT of the {\sc xcov}15 light curve after prewhitening
  (see text) by the 10 periodicities with the highest amplitudes:
  these frequencies are indicated by the downward (red) arrows.  The
  vertical axes use the same units as in plot 2 and the horizontal
  dashed lines at 0.54 mma in the plots correspond to the 0.001 false
  alarm probability detection threshhold established by the Monte
  Carlo data-shuffling method discussed in the text.  Using this value
  we can assert that there is real power at $\sim$ 4800 $\mu$Hz
  (f$_{9}$), but not at the frequency just below 4000 $\mu$Hz marked
  a.  The vertical (green) lines in the lower panel, annotated with
  the letters b - h, correspond to the predicted positions of
  combination frequencies where there are indications of real power in
  the DFT (see Table~\ref{tab:freq}).  The detected frequencies
  labelled f$_{15}$ and f$_{16}$ having amplitudes just above the
  significance threshhold also correspond to the predicted values of
  combination frequencies.}
\label{fig:wdft}
\end{figure*}

Direct visual inspection of the DFT in Fig.~\ref{fig:dft} allows one
to readily conclude that there are about fifteen `real' periodicities
present in the data.  To this end, the fifteen visually obvious
frequencies, ranging from $\sim 2000 \mu$Hz to $10^4 \mu$Hz, have been
labelled f$_1$ to f$_{18}$ in order in the plot, and also listed in
Table~\ref{tab:freq}.  At this stage a somewhat conservative (but
arbitrary) threshhold has been used to decide that the peaks f$_{13}$
and f$_{18}$, for example, are real and other smaller amplitude peaks
(such as f$_{9}$ and others at higher frequencies) are uncertain.
Note that the vertical amplitude scales in the upper and lower panels
of Fig.~\ref{fig:dft} are different, and that the reality of f$_{7}$
is clear from either a careful comparison of the DFT in the region of
f$_{8}$ and the form of the window function, or from the analysis
presented below.

The EC\,20058 light curve is dominated by the beating between the two
largest $\sim$ 8 mma pulsation modes at 281\,s and 257\,s (f$_6$ and
f$_8$, respectively).  With these modes combining to produce a maximum
modulation of less than 2\% ($\sim$ 17 mma) every 50 minutes, this
star is certainly not a large amplitude pulsator.  However, as noted
previously, the actual intrinsic stellar flux amplitudes are
approximately 40\% larger due to the uncorrected companion star
contamination.

\subsection{Light curve prewhitening}

A very useful procedure for separating peaks in the DFT corresponding
to real power from those peaks produced by window effects and/or noise
is that of `prewhitening'.  In this procedure, one or more selected
frequencies are first removed from the light curve by subtracting
least-squares fitted sinusoids, and then a DFT of the `prewhitened'
light curve is computed.  This method is particularly enlightening
when one is attempting to identify small amplitude periods that are
mixed in with the alias peaks of a much larger amplitude perodicity.
As pointed out by \citet{sca82}, least-squares fitting of sinusoids in
the time domain and computing DFTs are essentially identical
analytical procedures, so they both should independently lead to the
same conclusions.  Consequently, a careful comparison of the window
function with the DFT spectrum in the vicinity of a peak should
provide the same information as that revealed by the prewhitening
procedure.  However, uncertainties in the plots sometimes dominate
such comparisons, and therefore the prewhitening method is often more
decisive, as some examples below make clear.

\begin{table}
\caption{Detected frequencies in the DFTs for three EC\,20058 data
  sets.  The amplitudes (in millimodulation units) for the
  1997 {\sc xcov}15 WET run are given in column 5, while the
  corresponding values for the 2004 Mt John data and the SAAO 1994
  discovery data are given in columns 6 and 4, respectively.  The
  adopted combination frequencies are listed in the last column and
  amplitude values below the relevant significant levels are given in
  parentheses.  Note that the amplitude values have not been adjusted
  to account for the companion field star light contamination and the
  1994 values have been obtained from a subset of the Koen et
  al. (1995) data -- 4 runs in July 1994.}
\label{tab:freq}
\vspace*{5mm}
\begin{tabular}{@{}lrrcccl}
\hline
\multicolumn{1}{l}{item}   &
\multicolumn{1}{c}{freq.}  &
\multicolumn{1}{c}{period} &
\multicolumn{3}{c}{amplitude [mma]} &
\multicolumn{1}{c}{combn.} \\
\cline{4-6}
\multicolumn{1}{c}{}          &
\multicolumn{1}{c}{[$\mu$Hz]} &
\multicolumn{1}{c}{[s]}       &
\multicolumn{1}{c}{1994}   &
\multicolumn{1}{c}{1997}   &  
\multicolumn{1}{c}{2004}   &
\multicolumn{1}{c}{freqs.} \\ 
\hline

\ \ f$_1$     &  1852.6  & 539.8  &      & 2.0 &     &   \\
\ \ f$_2$     &  1903.5  & 525.4  &      & 1.9 & 1.5 &   \\
\ \ f$_3$     &  2852.4  & 350.6  & (1.6)& 1.4 & 1.3 &   \\
\ \ f$_4$     &  2998.7  & 333.5  & 2.2  & 3.0 & 2.6 &   \\
\ \ f$_5$     &  3489.0  & 286.6  &      & 1.7 & 1.3 &   \\
\ \ f$_6$     &  3559.0  & 281.0  &  8.8 & 8.5 & 7.6 &   \\
\ \ f$_7$     &  3640.1  & 274.7  &      & 1.0 &     &   \\
\ \ f$_8$     &  3893.2  & 256.9  &  8.0 & 8.4 & 7.4 &   \\
\ \ a         &  3924.2  & 254.8  &      & (0.5) &   &  \\
\ \ f$_9$     &  4816.8  & 207.6  &      & 0.7   &   &  \\
\ \ f$_{10}$  &  4887.8  & 204.6  & 4.2  & 2.7  & 2.4 &   \\
\ \ f$_{11}$  &  4902.2  & 204.0  &      & 1.7  & 1.4 & f$_2$+f$_4$\\
\ \ f$_{12}$  &  5128.6  & 195.0  & 3.5  & 2.5  & 1.8 &  \\
\ \ b         &  5462.5  & 183.1 &      & (0.5) &    & f$_2$+f$_6$ \\
\ \ c         &  5745.8  & 174.0 &      & (0.3) &    & f$_1$+f$_8$ \\
\ \ f$_{13}$  &  6557.6  & 152.5  &     & 0.6   &     &  f$_4$+f$_6$ \\
\ \ d         &  6745.6  & 148.2  &     & (0.5) &    &  f$_3$+f$_8$ \\
\ \ e         &  7118.0  & 140.5  &     & (0.5) &    &  2f$_6$      \\
\ \ f$_{14}$  &  7452.2  & 134.2  & 2.3 & 1.7  & 2.0  & f$_6$+f$_8$ \\
\ \ f         &  7533.5  & 132.7  &     & (0.4) &     & f$_7$+f$_8$ \\
\ \ g         &  8127.3  & 123.0  &     & (0.5) &     & f$_4$+f$_{12}$ \\
\ \ h         &  8446.8  & 118.4  &     & (0.5) &     & f$_6$+f$_{10}$ \\
\ \ f$_{15}$   &  8461.2 & 118.2  &     & 0.6   &     & f$_2$+f$_4$+f$_6$ \\
\ \ f$_{16}$   &  8781.0 & 113.9  &     & 0.6   &     & f$_8$+f$_{10}$ \\
\ \ f$_{17}$   &  9021.7 & 110.8  & 1.6 & 0.9  & (1.1)& f$_8$+f$_{12}$ \\
\ \ f$_{18}$   & 10030.7 & 99.7   &     & 0.7  &      & f$_2$+f$_4$+f$_{12}$ \\
\hline

\end{tabular}
\end{table}

Fig.~\ref{fig:wdft} shows the WET DFT covering the same frequency
range as for Fig.~\ref{fig:dft}, but with the light curve first
prewhitened by the 10 highest amplitude frequencies: f$_1$ -- f$_4$,
f$_6$, f$_8$, f$_{10}$ -- f$_{12}$ and f$_{14}$.  The y axis scales in this
plot have been expanded, and the removed periodicities are indicated
by the downward arrows.  The prewhitening was carried out by
simultaneously fitting (via least squares) 10 sinusoids to the data
using periods obtained from the original DFT.  If the periods are kept
fixed, linear least squares fitting can be used, and in practice this
is adequate for the well-defined frequencies in the data (especially
if accompanied by a limited grid search).

An examination of the Fig.~\ref{fig:wdft} DFT shows that there are two
clearly significant peaks (f$_5$ and f$_7$) on either side of the main
peak (f$_6$).  In addition, there are two peaks in the upper panel (a
and f$_{9}$) that look as though they could be significant, a
collection of 9 peaks in the lower panel (labelled b -- h, f$_{15}$,
f$_{16}$), many of of whom could be significant, and one or two peaks
at low frequencies that need to be investigated.  These latter peaks
will be discussed in the next section, but clearly we need a
quantitative criterion for distinguishing between real and noise
peaks.

\begin{figure*}
\includegraphics[width=6.5cm, angle=-90]{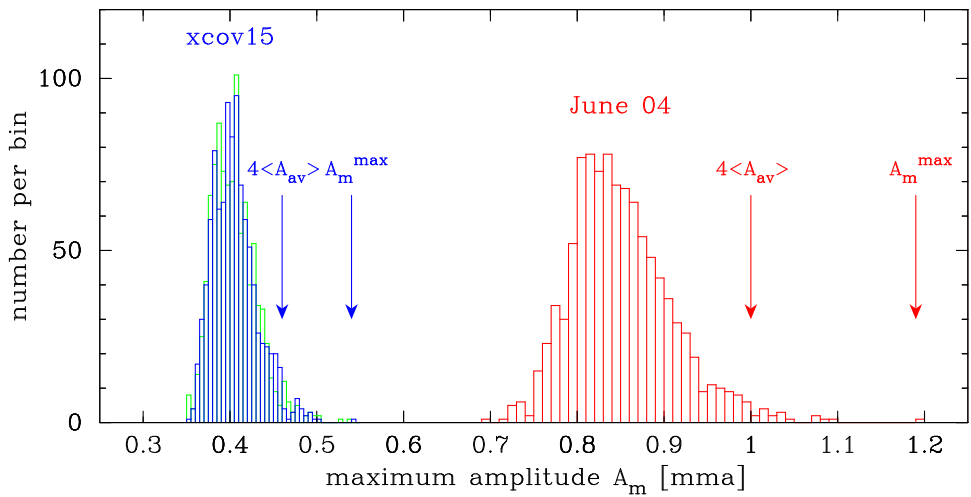}
\caption{Histograms showing the results of the Monte Carlo data
  shuffling exercises (see text) invoked to establish a false alarm
  probability detection threshhold (0.001) for real power in the
  prewhitened DFTs.  The horizontal axis gives the magnitude of the
  maximum peak in each DFT of the randomised time series data <and the
  vertical axis gives the number of occurrences per bin interval.  The
  two histograms (blue and green) centred around 0.4 mma correspond to
  separate 1000 shuffling trials of the pre-whitened {\sc xcov}15 data
  and use a bin interval of 0.005 mma, while the histogram on the
  right employs a bin interval of 0.01 and represents the same process
  applied to the 2004 multi-night Mt John data.  For each DFT of the
  randomised time series, the maximum peak A$_\mathrm{m}$ in the range
  1600 - 10,000 $\mu$Hz was selected.  The downward arrows mark the
  maximum peaks obtained in all 1000 trials (A$\mathrm{_m^{max}}$) and
  4 times the average DFT peak height (see text) for each ensemble of
  the DFTs ($4<$A$\mathrm{_{av}}>$).  See the online journal article
  for a colour version of this figure.}
\label{fig:hist}
\end{figure*}

A detailed inspection of the prewhitened DFT for the entire data set
listed in Table~\ref{tab:x15} in the vicinity of the dominant periods
f$_{6}$ and f$_{8}$ revealed that there was a small but significant
amount of power left in the form of residual `window mounds' for both
frequencies.  In other words, the subtraction of the two periodicities
from the light curve had not removed all of the power at those
frequencies.  For a pulsating white dwarf exhibiting very stable and
highly coherent luminosity variations, this can be caused by one or
more data segments having a timing error.  By a process of trial and
error, the `culprits' were determined to be the two OPD observation
files ra411 and ra413 marked with an asterisk in Table~\ref{tab:x15}.
Upon deletion of these two files from the overall light curve, the
prewhitened DFT produced the results graphed in Fig.~\ref{fig:wdft},
and shown in more detail in Fig.~\ref{fig:f6+f8} (left panels): no
signatures of the subtracted frequencies remain.  This discovery
illustrates the utility of the prewhitening procedure, as the
`contribution' from the two relatively small data segments could
easily have been overlooked otherwise.  In order to be certain, the
timing discrepancies were investigated graphically by overlaying a
time series plot of the two data segments with the predicted light
curve obtained from a least squares fit of the two dominant
frequencies to the rest of the WET data.  Both plots clearly revealed
the presence of a timing error.

Given the unlikely nature of the alternative hypothesis --- the star
was behaving differently during these periods --- subsequent analysis
removed these data from the combined WET light curve.  No obvious
explanation for the timing error has been found.

\subsection{DFT noise simulations}

A useful quantitative criterion for differentiating between signal and
noise peaks in the DFT is the concept of the false alarm probability
(FAP) introduced by \citet{sca82}, which makes the statistical nature
of the process explicit.  Using this concept, one computes the
probability of a given peak in the DFT being due to noise for some
chosen threshhold.  A high threshhold for false positives can be set
by ensuring that this probability is small, and only counting peaks
that are above this threshhold as real.

Although the FAP can be estimated theoretically on the basis of some
assumed model for the noise characteristics \citep[e.g.][]{sca82}, it is
preferable and more representative of the actual data to determine it
by a Monte Carlo simulation method.  And, assisted by the speed of
modern computers, this is a practical proposition for even large data
sets.  Such a simulation has been carried out for the data sets
discussed here using the program
\emph{tsmran}\footnote{http://whitedwarf.org/tsmran} developed by one
of us (DJS).  The results of this work for the WET data set are that
peaks with an amplitude of 0.54 mma have a FAP value of 0.001, meaning
that there is \emph{only} one chance in 1000 that such a peak can be
produced by a random noise `conspiracy'.  The corresponding value for
the Mt John 2004 (MJ04) data set is 1.2 mma.

\begin{figure*}
\includegraphics[width=11.5cm, angle=-90]{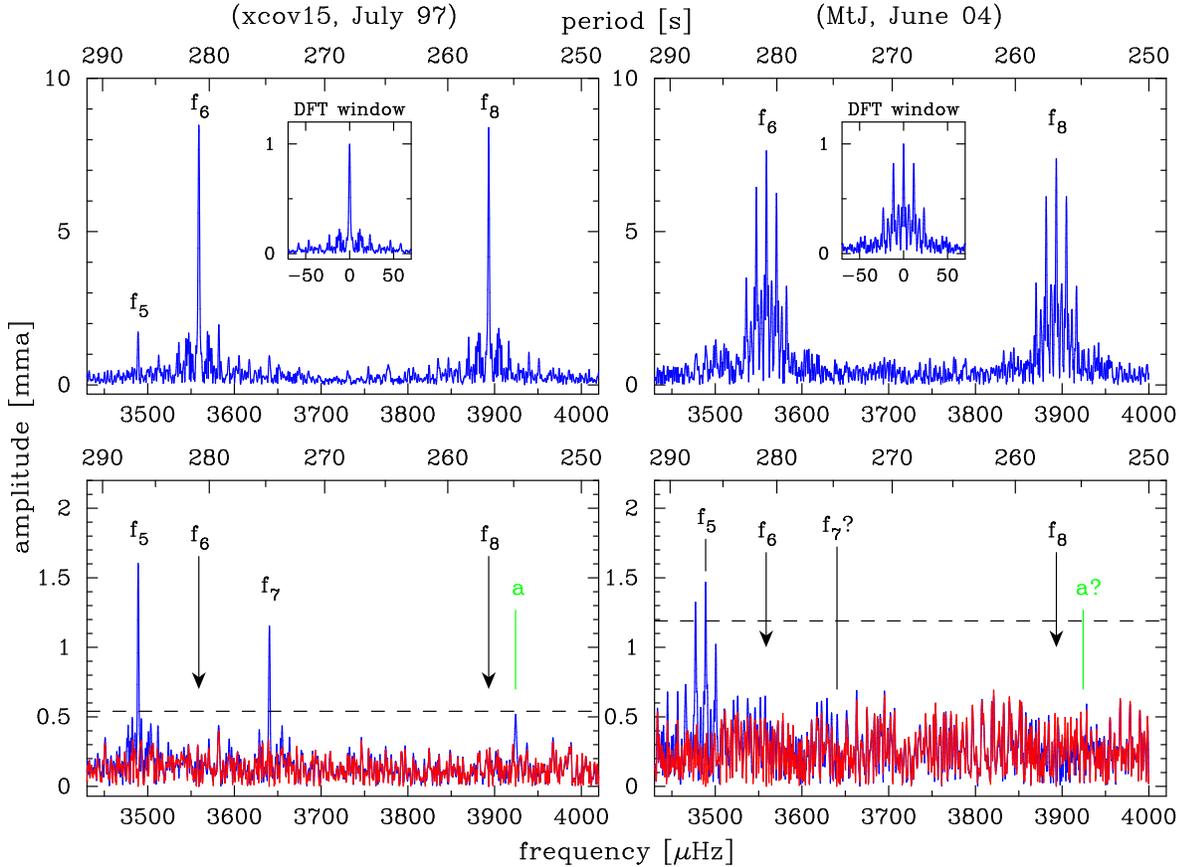}
\caption{Expanded amplitude DFTs of the frequency region near modes
  f$_6$ and f$_8$, comparing the 1997 {\sc xcov}15 data set (top left) with
  the June 2004 (Mt John) data set (top right).  The panels at the
  bottom, using an expanded amplitude scale, depict the corresponding
  DFTs of each time series prewhitened by the frequencies f$_6$ and
  f$_8$ (blue curves). The periodicities corresponding to f$_5$ and
  f$_7$ are clearly evident in the prewhitened {\sc xcov}15 data
  (bottom left) but only f$_5$ is evident in the Mt John data.
  There is a suggestion of power (marked `$a$') in the {\sc xcov}15 data
  adjacent to f$_8$ that is just below the adopted detection
  threshhold (dashed line).  Both bottom panels also include overlay
  plots (in red) of prewhitened DFTs in which the remaining
  periodicties have been removed.  See the online journal article for
  a colour version of this figure.}
\label{fig:f6+f8}
\end{figure*}

\begin{figure*}
\includegraphics[width=11.5cm, angle=-90]{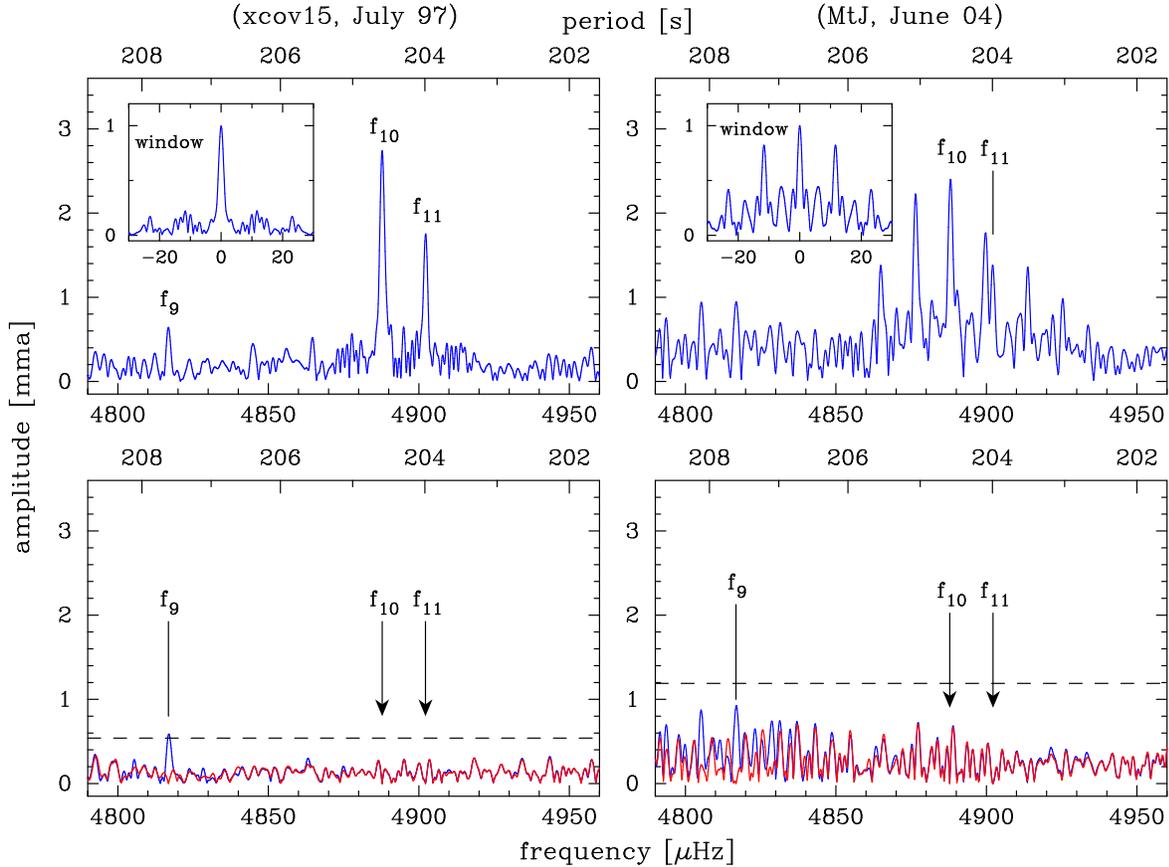}
\caption{Expanded amplitude DFTs of the frequency region near modes
  f$_{10}$ and f$_{11}$, comparing the 1997 {\sc xcov}15 data set (top left)
  with the June 2004 (Mt John) data set (top right).  The panels at
  the bottom depict the corresponding DFTs (blue curves) of each time
  series prewhitened by the frequencies f$_{10}$ and f$_{11}$.  The
  horizontal dashed lines in the bottom panels represent the detection
  threshholds for significant power in the respective data sets (see
  discussion in Section 4.3).  The f$_{9}$ frequency is above the
  detection threshhold in the WET data and, although it is below the
  Mt John data detection threshhold, it does appear to make its
  presence felt.  The red overlay plots in both bottom panels
  represent DFTs of each data set further prewhitened by the f$_{9}$
  periodicity.  See the online journal article for a colour version of
  this figure.}
\label{fig:f10+f11}
\end{figure*}

These threshholds are marked as horizontal dashed lines in appropriate
parts of Figs.~\ref{fig:wdft}, \ref{fig:f6+f8} and \ref{fig:f10+f11}
showing the prewhitened DFTs for both data sets.

Although the simulation procedure is conceptually straight forward,
there are several points that should be noted and a number of
organisational tools are required for a flexible implementation.  So,
we briefly outline the \emph{tsmran} methodology here.

The essential idea is to start with the reduced light curve that has
been prewhitened by the `clearly real' frequencies, randomly
rearrange the time order of the data points, compute a DFT of these
randomised data covering the frequency range of interest, and then
record both the maximum (A$_\mathrm{m}$) and the average
(A$_{\mathrm{av}}$) value of peak heights in the DFT.  Randomising the
time order of the data destroys the coherency of any periodic signal
remaining in the light curve but preserves the uncorrelated noise
characteristics of the data.  The highest peak in the DFT of a
time-shuffled data set is then a one sample estimate of the maximum
excursion that can occur due simply to random noise effects. If this
procedure is repeated a large number of times (say 1000), and the
highest peak in each DFT recorded, then one can infer that the maximum
value, A$_{\mathrm{m}}^{\mathrm{max}}$, in this ensemble of highest
peak values is a direct estimate of the DFT amplitude threshhold for a
false alarm probability at the 1 in 1000 level.

It is instructive to plot a histogram of the recorded maximum peaks
(A$_\mathrm{m}$) for each DFT: 1000 member ensembles for both the WET
and MJ04 data are depicted in Fig.~\ref{fig:hist}.  There are two
separate such histograms (blue and green) for the WET data centred
around 0.4 mma, and one for the MJ04 data centred around 0.9 mma.  It
is clear that the less comprehensive MJ04 data exhibits larger
A$_\mathrm{m}$ values, as one would expect.

The program \emph{tsmran} also records the average peak height for
each randomised DFT, partly as a check on its operation, and as one
might expect there is little variation from DFT to DFT.  The
histograms of these average peak height values have not been shown,
but four times the ensemble average ($4<$A$_{\mathrm{av}}>$) for each
data set has been indicated in Fig.~\ref{fig:hist}, as one sometimes
sees a criterion similar to this used in the literature to identify
real power \citep[e.g.][]{kep93}.  Note, however, that we are using
here (the average of) the average amplitude DFT values and \emph{not}
the square root of the average power spectrum values.

A few more comments about the algorithms employed in \emph{tsmran} are
relevant.  For the relatively large data sets considered here (WET
$\sim$ 45k and MJ04 $\sim$ 15k integrations, respectively) a long
period random number generator is required: the routine \emph{ran2}
from the Numerical Recipes suite \citep{pre92} was found to be equal
to the task.  One also makes significant efficiency gains by using a
fast Fourier transform (FFT) algorithm to compute the DFTs for
suitable data sets; the routine \emph{realft} from \citeauthor{pre92}
accomplishes this task in \emph{tsmran}.

However, since FFT algorithms assume that the time-series data
correspond to equal contiguous sample intervals, some preliminary
organisation of the data segments is required before a FFT can be
employed.  Since all the observations reported in this paper
correspond to 10\,s integrations it was possible to create an array of
integrations that satisfied this attribute (with adequate precision
for the intended purpose) without resorting to such (undesirable)
procedures as interpolation.  All deleted integrations in each
`significant' data segment were restored using zero padding, and the
shape of the overall DFT window function was essentially preserved by
separately randomising the points in each segment before combining
them together in one overall time series, all the while maintaining
the data gaps using zero padding.  Last, the combined time series is
extended to an integer power of two (as required by \emph{realft})
using zero padding, in order to maximise the efficiency of the FFT
procedure.

A comment on the data timescale is appropriate.  Although the above
assumption of uniform contiguous integration intervals represents the
original terrestrial TAI observation timecale, offsets are introduced
after the transformations to the BJD timescale.  This will have little
impact on use of the FFT for the simulation exercise, and negligible
impact anyhow if the offsets are not very large.  For the record, the
total offset change across the WET data is $\sim$ 7\,s, and for the
MJ04 data it is $\sim$ 25\,s.

Individual data segments corresponding to each run are read separately
into \emph{tsmran} using a list of file names in an input file
provided by the user.  A glance at Fig.~\ref{fig:ts} (especially the
colour version) shows that there is some data segment overlap (eg, the
fourth panel from the top: SAAO, CTIO and MJUO on 6 July).  These
overlaps were removed by deleting the minimum number of points in
order to allow contiguous data to be created.

A version of the program exists which uses the much slower
(computationally inefficient) direct DFT algorithm.  This then
obviates the need for much of the data organisation code in
\emph{tsmran}, which admittedly was a significant proportion of the
programming coding effort.  This method is more general and can be
invoked for non-uniform time series data sets, but it is
\emph{significantly} slower.  In contrast, the FFT method is extremely
fast: on a `standard' modern laptop the one thousand 64k WET DFTs
are computed in less than a minute --- effectively `on the fly', one
might say.

At this point it is relevant to emphasise that our simulations are
dealing with a random noise model only; other non-random sources of
`noise' (eg, periodic telescope drive error) need to be handled on a
case-by-case basis.  In addition, we limited our region of interest to
frequencies above 1600$\mu$Hz, as (a) there are no obvious peaks in
the DFT in this region that stand out above the forest of other peaks
and (b) we would need to consider a higher detection threshhold due to
the increasing impact of residual sky noise.

\subsection{Detected frequencies}

Referring to Figs.~\ref{fig:dft}, \ref{fig:wdft}, \ref{fig:f6+f8} and
\ref{fig:f10+f11}, and Table~\ref{tab:freq}, we have identified 18
definite frequencies (f$_*$) in the WET data as being real, using the
adopted 0.001 false alarm probability.  The amplitudes for these
frequencies are given in column 5 of the table, while the amplitudes
of other possible frequency detections (a -- h) are listed in the
table within parentheses.

Two frequencies that merit further discussion are `$a$' and and
f$_{9}$; they are depicted in more detail in the panels of
Figs.~\ref{fig:f6+f8} and \ref{fig:f10+f11} and .  These frequencies are
essentially outside the region where the issue of linear combinations
arises (see next section) and their amplitudes are very close to the
0.54 mma significance threshhold.  They only first appear as
interesting in the prehwhitened DFT of Fig~\ref{fig:wdft}, and the
MJ04 data set provides additional information.  This is depicted in
the right panels of Figs.~\ref{fig:f6+f8} and \ref{fig:f10+f11} by
graphing both the MJ04 DFTs and prewhitened DFTs covering the same
frequency ranges as for the WET DFTs.

The {\sc xcov}15 case for f$_{9}$ is clearly presented in the two
left panels of Fig.~\ref{fig:f10+f11}.  The top left panel shows the
DFT with this frequency marked (along with f$_{10}$ and f$_{11}$), the
lower left panel (blue curve) shows the DFT prewhitened by f$_{10}$ and
f$_{11}$ and power at the f$_{9}$ frequency above the 0.54 mma
threshhold, and the dark (red) curve shows the DFT further prewhitened by the
f$_{9}$ frequency in which the peak has disappeared.

The two right panels in Fig.~\ref{fig:f10+f11} present the MJ04 data
case in the same format as the left panels.  There is clearly evidence
of power in these data at the frequency corresponding to f$_{9}$ which
follows the shape of the window function and whose maximum peak height
is not far below the relevant 1.2 mma significance threshhold.  Also,
the further prewhitening procedure removes these peaks from the DFT
(red curve).  Although the principal peak is below the 1.2 mma MJ04
data significance threshhold, so in isolation would not be interpreted
as a detection using our stated threshhold, it does provide definite
supporting evidence for the {\sc xcov}15 detection.

On the other hand, the case for (and against) `$a$' is presented in
a similar manner in Fig.~\ref{fig:f6+f8}.  In the lower left panel,
the blue curve depicts the {\sc xcov}15 data prewhitened by the two
dominant frequencies (f$_{6}$ and f$_8$) and the expanded amplitude
scale clearly shows the reality of f$_5$ and f$_7$, and also a
possibly real frequency `$a$', tantalizingly just below the 0.54
mma significance level.  The dark (red) curve is a DFT further
prewhitened by these three frequencies and emphasises the point that
the `$a$' peak could represent real power.  However, the same
procedure for the MJ04 data tells a different story.  The blue curve
in the lower right panel shows the MJ04 DFT prewhitened by the
dominant f$_{6}$ and f$_8$ periodicities and there is no evidence of
`$a$'.

It is interesting that there is also no evidence for f$_7$
in the MJ04 data.  However, there is without question power at this
frequency in the WET data so further confirmation is not required.  It
does illustrate, though, that power in this quite stable pulsator at
various frequencies is still variable at a low level, the result of
real physical amplitude instabilities, or perhaps beating.

In view of the above, we have included f$_9$ as a real detected
periodicity, but not included `$a$' --- we leave it in the summary
table as a bracketed `maybe'.

\subsection{Linear combination frequencies}

Since the mechanism that converts mechanical movement of stellar
material to luminosity variations at the surface appears to be
nonlinear, both harmonics and combinations of the basic mode
frequencies can be expected to appear in the light curve.  At least,
this is the conclusion of many previous studies of the pulsating white
dwarfs --- especially the large amplitude pulsators, such as DBV
GD\,358 \citep{kep03}.

The last column in Table~\ref{tab:freq} lists whether any of the
detected frequencies in the light curve can be identified as a linear
combination of other detected frequencies, as well as identifying
whether possibe linear combinations could account for
`suspiciously' large peaks that are below the 0.54 mma detection
threshhold; all these quantities are also indicated on the plot in
Fig.~\ref{fig:wdft}.

So, all of the actual 7 detected frequencies f$_{11}$ and f$_{13}$ --
f$_{18}$ can be explained as combination frequencies.  This
interpretation leaves f$_{12}$ as the highest frequency pulsation
mode.

The two largest combination frequencies with amplitudes $\sim$ 1.7 mma
merit further discussion.  The frequency f$_{14}$, being simply the sum
frequency of the two highest amplitude modes (f$_{6}$ and f$_{8}$)
is obviously the first combination to look for, but f$_{11}$ being
considered a combination of the lower amplitude modes f$_{2}$ (1.9
mma) and f$_{4}$ (3.0 mma) is perhaps surprising.  Furthermore, these
two modes appear to combine separately with f$_{6}$ and f$_{12}$ to produce
signal power at f$_{15}$ and f$_{18}$, respectively.

\begin{figure*}
\includegraphics[width=12.0cm, angle=0]{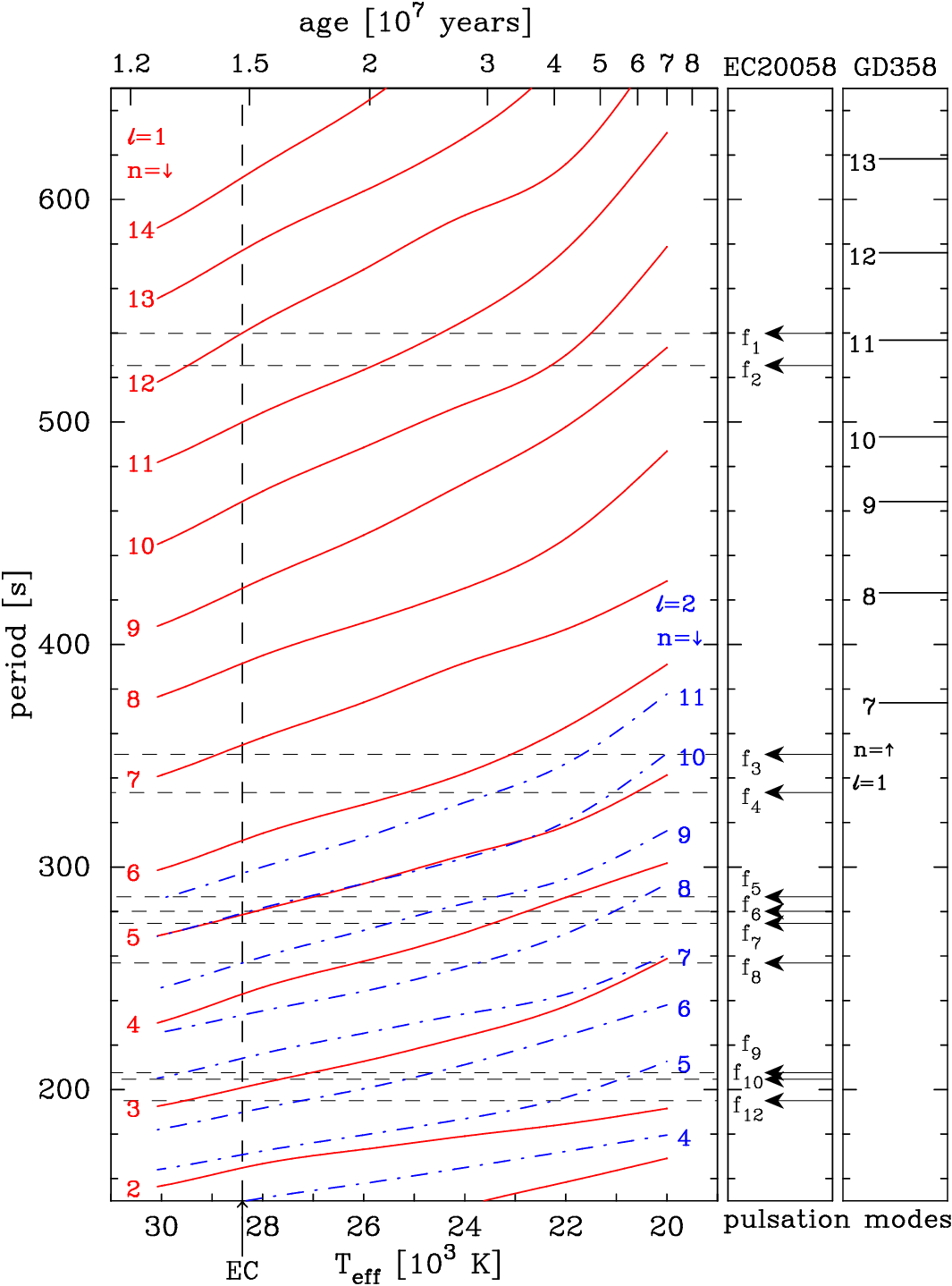}
\caption{A pulsation mode diagnostic diagram showing the predicted
  periods of the pulsation modes as a function of effective
  temperature for a range of 0.6M$_{\sun}$ white dwarf models.  The
  left panel graphs all $\ell = 1$ modes (red solid lines) in the
  period range and only some of the $\ell = 2$ modes (blue dash-dot
  lines).  The middle panel shows the detected modes for EC\,20058
  and the right panel shows the modes detected for GD\,358 for
  comparison purposes.  See text for more details and the online
  journal article for a colour version of this figure.}
\label{fig:model}
\end{figure*}

Whereas some of the higher frequency combinations would be difficult
to explain as direct mode frequencies using realistic white dwarf
models, this is not true for the frequency f$_{11}$: it is not
absolutely certain that this frequency results from nonlinear
combination effects.  However, given the frequency matches mentioned
above, and the likely difficulty of explaining the presence of the
four closely spaced frequencies f$_9$ -- f$_{12}$ in terms of
low-order $\ell$ pulsation modes (see Fig.~\ref{fig:model} and
discussion next section), we adopt the conservative approach and
delete f$_{11}$ from the list of inferred modes.

Column 5 in Table~\ref{tab:freq} lists the 11 frequencies (f$_{1}$ --
f$_{10}$ and f$_{12}$) that we consider represent pulsation modes from
the work presented here, while columns 4 and 6 in the same Table list
for comparison all the frequencies detected in the work of
\citet{koe95} and independently in the MJ04 data set, respectively.

\section{Comparison with models}

Our aim here is to compare the 11 detected pulsation modes for
EC\,20058 with predictions from white dwarf models in order to tell us
something about the star.  There are two basic approaches to this
task.  In the first instance we can look for relatively simple
systematic trends in the observed pulsation spectrum that depend on
stellar properties.  There are, in fact, two of these: one is a
sequence of periods with approximately equal spacing (which is
obviously best viewed in period space), while the other is
`mode-splitting', which is seen more clearly in frequency space.
Second, we can adopt a global fitting procedure which endeavours to
match the predicted modes to the data, such as that pioneered by
\citet*{met00}.

We need two basic results from the relevant pulsation theory in order
to look for any systematic trends.  The buoyancy-driven g-mode
pulsations in white dwarfs \citep[e.g.][]{unn89,han04} can be
characterised by three integer quantum indices: $n$ (or sometimes
$k$), $\ell$ and $m$.  These indices specify a unique eigenmode of
oscillation in which $n$ determines the radial form of the
eigenfunction and $\ell$ and $m$ specify the angular behaviour via a
spherical harmonic function $Y_{\ell \mathrm{m}}(\theta,\phi)$.  If
spherical symmetry holds (no rotation and/or magnetic field), then the
mode periods, $\Pi_{n \ell}$, are independent of the index $m$ and in
the limit of large $n$ we can write
\begin{center}
$$
    \Pi_{n \ell} \approx \Pi_{0} \frac{n + \epsilon}{\sqrt{\ell(\ell + 1)}}
     \hspace{4mm} \Longrightarrow \hspace{4mm}
    < \! \Delta \Pi_{\ell}\!> \: \approx \, \frac{\Pi_{0}}{\sqrt{\ell(\ell + 1)}}
$$
\end{center}
The quantity $\Pi_{0}$ is inversely proportional to an integral of the
Brunt-V\"{a}is\"{a}l\"{a} (or buoyancy) frequency over the star
profile, and hence decreases with increasing stellar mass (due to
increasing local gravitational field strengths). Hence, $\Pi_{0}$
establishes a basic timescale for the period structure; $\epsilon$ can
be neglected for our purposes.

Rotation of the star removes the
$m$-degeneracy of the mode frequencies, and to \emph{first order}
leads to mode frequencies that are a function of all three quantum
indices in the form
\begin{center}
$$
 \nu_{n \ell m} = \nu_{n \ell} + m(1 - C_{n \ell}) \Omega
$$
\end{center}
where $\nu_{n \ell}$ is the degenerate frequency, $m$ takes the $2
\ell + 1$ integer values between $-\ell$ and $+\ell$, $\Omega$ is the
rotation frequency of the star and $C_{n \ell}$ takes the approximate
dimensionless form $C_{n \ell} = 1/\ell(\ell + 1)$ for high radial
overtone modes.  The net effect of this is that all ($n$,$\ell$)
g-mode pulsations undergo a `Zeeman-like' splitting due to the
rotation of the star which can theoretically be seen as 2$\ell$+1
closely spaced frequencies (see \citealt{cox84} for a nice physical
explanation).

A magnetic field also destroys the spherical symmetry and leads to
frequency splitting: in the case of a small magnetic field aligned
with the pulsation axis, $\ell+1$ splitting occurs \citep{jon89}.
Whether any of these modes are sufficiently excited such that they are
detectable is another story.

\subsection{Model calculations}

In order to provide concrete examples, we have determined a range of
pulsation modes for a sequence of WD models with effective
temperatures between 30\,000 and 20\,000 K.  The results of our
calculations are displayed in Fig.~\ref{fig:model} (left panel), along
with the detected mode periods for EC\,20058 (middle panel) and the
DBV class protype (GD\,358, right panel).

The WD models and the respective pulsation frequencies were calculated
using code that has itself evolved over the years, and is described in
a succession of PhD theses at the University of Texas, Austin -- see
\citet{met00} for a brief description and appropriate references.  For
our fiducial model, we chose a mass of 0.6\,M$_{\odot}$, a 50/50 C/O
core and a fractional helium layer mass of $10^{-3}$.  Six evolved WD
models with effective temperatures at 2000\,K intervals between
30\,000\,K and 20\,000\,K were computed, and the pulsation modes
determined for each of these.  The pulsation modes for models of
intermediate temperatures between these values were estimated by
fitting cubic splines to all the mode periods as a function of
effective temperature and the results are plotted in
Fig.~\ref{fig:model}: all the $\ell=1$ mode periods are represented by
solid (red) lines in the displayed period range, and a few of the
shorter period $\ell=2$ modes are displayed using the (blue) dash-dot
lines.  Since we are using these model calculations to aid the
phenomenology discussion below, the precise values of the model
parameters are unimportant.

A very obvious feature of the pulsation period structure in the WD
models is that all mode periods increase with decreasing effective
temperature \citep*[e.g.][]{bra93}, and therefore increasing age.
Over human timescales this effect is very small, but nevertheless the
extremely high stability of many white dwarf pulsators means that this
effect can be observed \citep[e.g.][]{kep05a}.  We will discuss the
prospects for EC\,20058 in the last section.

We have chosen to discuss only the lowest order $\ell=1$ and $\ell=2$
pulsation modes, as we make the usual assumption \citep{dzi77} that
the effect of geometrical flux cancellation across the observable
stellar disk should ensure that modes with larger $\ell$ values, if
excited, contribute little to the observable flux changes.  However,
at this point it is appropriate to sound a small note of caution by
mentioning that there is evidence of a pulsation mode in the DAV
pulsator PY\,Vul (G\,185-32) that has an $\ell$ value of at least 3
and possibly 4 \citep{tho04,yea05}.  It is worth noting, though, that
this object exhibits an unusual pulsation spectrum.

For the discussion in the next section we will focus on some details
of the 28\,000\,K model, since this is closest to the effective
temperature obtained for EC\,20058 by \citet{bea99}, but see also
\citet{sul07}.  The pulsation modes in Fig.~\ref{fig:model} for this
model exhibit a mean period spacing of 37.4\,s and a range of 33.4 --
43.0\,s for the $\ell=1$ modes, and a mean value of 20.4\,s with a
range of 16.6 -- 23.5\,s for the $\ell=2$ modes.

Among other things, these model values demonstrate that the pulsation
theory summarised above does only predict \emph{approximately}
constant period spacings for a sequence of modes with a given $\ell$
and varying $n$.  Also, this theory predicts that the ratio of the
`constant' period spacings for sequences of $\ell=1$ and $\ell=2$ in
the star should equal $\sqrt{2(2+1)}/\sqrt{1(1+1)} = 1.73$.  However,
this ratio for the mean period spacings in our computed model
(covering the period range in Fig.~\ref{fig:model}) is 37.4/20.4 =
1.83.  This is close to the approximate asymptotic theory value, but
not identical, as we are in the low $n$ regime.

\subsection{Period and frequency phenomenology}

It follows from the previous discussion that in the first instance we
should look for a sequence of observed periods with approximately
equal spacing in the pulsation spectrum of EC\,20058.  As is clear
from Fig.~\ref{fig:model}, such a pattern was detected in GD\,358
\citep{win94,kep03}, and they were all readily interpreted as having
spherical degree $\ell=1$ due to the fact they displayed clear triplet
(rotational) splitting.  This conclusion was possible, inspite of the
complicated nature of the GD\,358 pulsation behaviour.

No such simple pattern is immediately apparent in the EC\,20058
spectrum.  Nevertheless, we will attempt to identify any
trends.  If one assumes that a number of modes in some sequence are
not excited above an observable threshhold, then one could consider
some of the pairs f$_1$--f$_2$ ($\Delta P = 14.4$\,s), f$_3$--f$_4$
(17.1\,s), f$_6$--f$_8$ (24.1\,s) [or f$_7$--f$_8$ (17.8\,s)] and
f$_{9}$--f$_{12}$ (12.6\,s) as visible members of such a sequence.
But even ignoring the close spacing between the pairs, the inter-pair
spacings do not match any assumed reasonable sequence.  Only the two
pairs f$_3$--f$_4$ and f$_7$--f$_8$ show a similar period spacing with
a mean value of about 17.5\,s.  However, the period spacing between
these pairs is not even close to being an integral multiple of this
mean value.  Even if it was, we would then face the task of
interpreting the small period spacing in terms of either a sequence of
only $\ell=2$ excited modes and/or an improbably large model mass
$\sim 0.8M_{\odot}$ (as period spacing decreases with model mass).

Another possibility is to temporarily put aside the question of the
model mass and consider the two pairs f$_6$--f$_8$ (24.1\,s) and
f$_1$--f$_2$ (14.4\,s) as members of separate $\ell=1$ and $\ell=2$
sequences, respectively.  The period spacing ratio is then 24.1/14.4 =
1.67, and this is consistent with both the predicted pulsation theory
value (1.73) and the actual model estimates given in the previous
section.  However, in addition to concerns about the implied total
model mass, it is hard to argue convincingly that only two pairs of
frequencies are clear evidence of the predicted sequence.  Also, we have
not independently established any $\ell$ values for the modes, using
for example rotational splitting, as was successfully exploited for
GD\,358.

The most likely scenario is that we are seeing a combination of
$\ell=1$ and $\ell=2$ excited modes, coupled with possible rotational
frequency splitting.  An inevitable conclusion is that the two modes
with the largest amplitude (f$_6 \sim$ 281\,s and f$_8 \sim$ 257\,s)
correspond to different $\ell$ values, presumably $\ell=1$ and
$\ell=2$ in some order.

We now investigate the possibility of rotational splitting of
pulsation modes in the frequency spectrum.  A qualitative inspection
of Figs.~\ref{fig:dft} and \ref{fig:wdft} suggests that some of the
modes $f_5$, $f_7$, $f_{9}$ and even the `not accepted' frequency
`$a$' might be caused by rotational splitting.  Using the $\Delta
f$ ($\mu$Hz) values $f_6 - f_5 = 70\,\mu$Hz, $f_7 - f_6 = 81\,\mu$Hz,
$a - f_8 = 31\,\mu$Hz and $f_{10} - f_9 = 71\,\mu$Hz, we find
that the first and fourth of these pairs lead to similar splitting
values of $\sim$ 70 $\mu$Hz.  If these phenomena are due to splitting,
and the modes in question (f$_6 \sim$ 281\,s and f$_{10} \sim$ 205\,s)
have $\ell = 1$, then the white dwarf is rotating with a period of
close to 2 hours.  We are of course assuming that only one non zero
$m$ component is excited to an observable amplitude out of a possible
two for each dipole mode.  Note that this interpretation also provides
further support (if any is needed) for our argument that real power is
present at the frequency f$_9$.  A less likely $\ell = 2$
interpretation for both modes would requires only one out of 4 possible
non zero $m$ value modes in each case to be excited, and would predict
a rotation period close to 4 hours.

\begin{figure}
\includegraphics[width=7.0cm, angle=0]{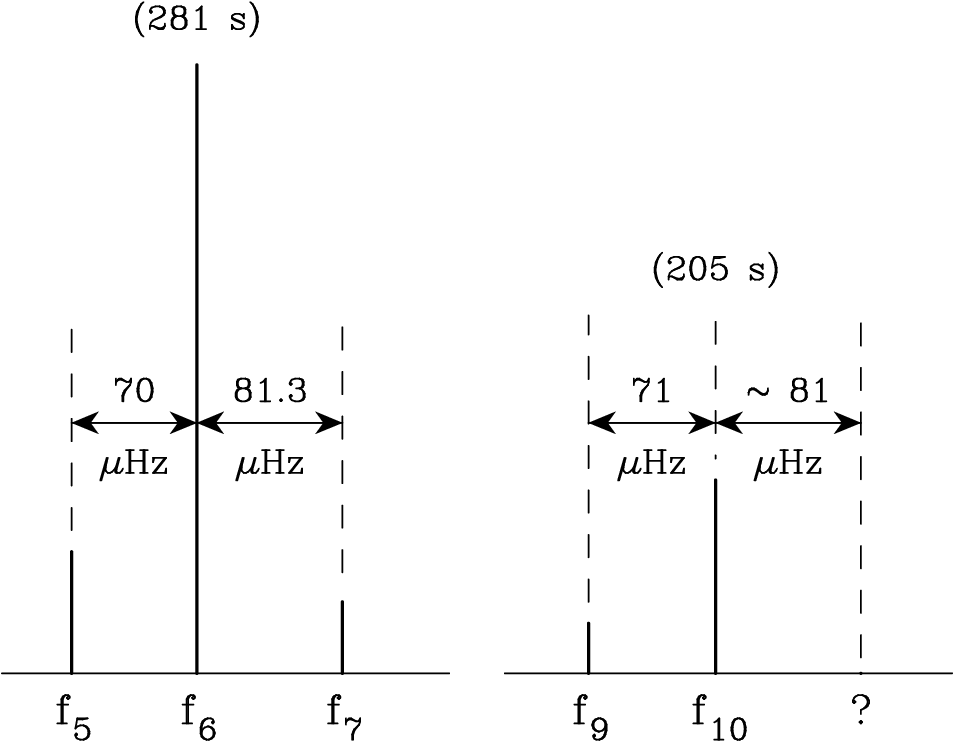}
\caption{A schematic amplitude spectrum illustrating the proposed
  rotational and magnetic mode splitting model.}
\label{fig:rotn}
\end{figure}

Given the deductions of other white dwarf rotation rates, a $\sim 2$ hour
period represents a rapidly rotating white dwarf.  This rotation rate
is certainly not unphysical, so we tentatively put it forward as a
possible interpretation, although the evidence is hardly overwhelming.
Note that the pre-white dwarf PG\,2131+066 has a measured rotation
rate of $\sim$ 5 hours \citep{kaw95}.  Interestingly, the issue of
measured white dwarf rotation rates is somewhat controversial as large
angular momentum losses via some mechanism are required in order to
make these measured rates consistent with the known rotation rates of
their much larger progenitors.

Before proceeding to the global modelling analysis presented in the
next section, we will include the small (1\,mma) satellite frequency
f$_7$ in the mode splitting model that we adopt.  A glance at
Fig.\ \ref{fig:f6+f8}, or the schematic version in Fig.~\ref{fig:rotn},
reveals that f$_5$ and f$_7$ appear to form an asymmetric pair either
side of the large amplitude `parent' mode f$_6$ ($\sim$ 281\,s) with
separations of 70\,s and 81.3\,s, respectively.  There are two ways to
produce asymmetric splitting: second-order rotation effects and the
inclusion of a magnetic field.

Second order rotation effects \citep[e.g.][]{chl78} have been detected
in the pulsation spectrum of the DAV pulsator L\,19$-$2
\citep{odo82,sul98}.  The dominant 192\,s mode for this pulsator
exhibits two low amplitude satellite modes that are separated by
$\sim$ 13\,$\mu$Hz from the main peak, and a rotational splitting
explanation yields a rotation period of $\sim$ 10 hours, assuming an
$\ell = 1$ value for the mode.  Furthermore, the frequency splitting
shows a 1\% asymmetry, which is consistent with a second order
rotational effect.  Even though EC\,20058 is estimated to rotate five
times faster and the second order effect depends on the square of the
rotation frequency, this is not enough to explain the asymmetry
displayed in Fig.~\ref{fig:rotn}.

The combined effect of rotation and a magnetic field will also produce
asymmetrical splitting.  Thus, if the frequency separations adjacent
to the f$_6$ (281\,s) mode are divided into a $\sim$ 75\,$\mu$Hz
symmetrical rotational splitting and a magnetic field frequency shift
(increase) of $\sim 5\,\mu$Hz, then this would explain the observed
triplet.  A magnetic field of $\sim$ 3\,kG and the 2\,h rotation
period would accomplish this \citep[][appendix]{win94}.  As suggested in
Fig.~\ref{fig:rotn}, evidence of a similar triplet structure around
the f$_{10}$ mode would add considerable weight to this proposal, but as
there is no evidence of power at f$_{10}+ \sim 81\,\mu$Hz, we simply
adopt these implied constraints in preparation for the global
modelling presented below.

\begin{table}
\caption{Optimal model parameters for EC\,20058.}
\label{tab:param}
\begin{center}
\begin{tabular}{lrrrr}
\hline
Parameter & 
\multicolumn{1}{c}{C} & 
\multicolumn{1}{c}{C/O} & 
\multicolumn{1}{c}{O} & 
\multicolumn{1}{c}{Average} \\
          & 
\multicolumn{1}{c}{(pure)} & 
\multicolumn{1}{c}{50:50}  & 
\multicolumn{1}{c}{(pure)} &  \\
\hline
$T_{\rm eff}\,(K)\dotfill$ & 28\,100  & 28\,200 & 28\,400 &$28\,250\pm150$ \\
$M_{\star}\ (M_{\odot})\dotfill$ & 0.550 & 0.550 & 0.550   &$0.550\pm0.005$ \\
$\log(M_{\rm env}/M_{\star})$   & $-$3.56 & $-$3.60 & $-$3.62 &$-3.59\pm0.03$ \\
$\log(M_{\rm He}/M_{\star})\dotfill$ & $-$6.42 & $-$6.46 & $-$6.48 &$-6.45\pm0.03$ \\
$\sigma_P\,({\rm sec})\dotfill$  & 1.89 & 1.94 & 2.03   &$\cdots$ \\
\hline
\end{tabular}
\end{center}
\end{table}

\subsection{Model period fitting}

Starting with a list of detected pulsation modes and then endeavouring
to deduce the parameters of a specific stellar model that reproduces
these modes is a classic example of an inversion problem in physics.
These tasks are notoriously difficult and are plagued by the lack of
unique solutions.  However, the apparent relative simplicity of a
white dwarf means that a realistic model can be specified by only a
few parameters and the potential richness of the $g$-mode pulsation
spectrum provides the possibility of many revealed modes, and hence
many constraints to restrict the inversion process.

Following a WET run on the object GD\,358 \citep{win94} that yielded
154 h of near continuous time series photometry, \citet{bra94} were
able to use the detected pulsation modes to identify a preferred model
for this star.  They were aided in this task by the fact that the 11
detected normal modes clearly formed a sequence of increasing radial
order ($n$) that were consistent with an $\ell = 1$ assignment, due to
the evidence of a triplet structure for each mode. Their asteroseismic
analysis enabled a determination of such model parameters as: total
mass, surface helium layer mass and effective temperature.  Luminosity
and a bolometric correction were also estimated, which then led to an
asteroseismic distance determination.

A further advance in these modelling procedures has been made more
recently by the use of a genetic algorithm to efficiently explore the
multidimensional parameter space in a global search for the optimum
model or models that fit the pulsation data \citep*{met00}.
\citet{met03a} (and references therein) reports an analysis of GD\,358
data that is sensitive to the presumed white dwarf's $^{12}$C/$^{16}$O
core composition, and thereby yields an indirect measurement of the
$^{12}${C}($\alpha$,$\gamma$)$^{16}${O} reaction cross section at
astrophysically relevant energies (as the C/O core composition is
determined by the competition between this reaction and the C-forming
triple $\alpha$ reaction).  Furthermore, \citet{met05} have
successfully exploited the global optimization method to analyse
pulsation data obtained from dual-site observations (and previous
work) of another DBV star CBS\,114, and clarify a number of features
of its structure.

Note that there has been some controversy concerning these
asteroseismic successes, which is related to the fact that both
composition variation in the core \emph{and} in the envelope below the
helium atmosphere produce similar non uniformities in the period
spectrum \citep*{fon02,bras03,mon03}.

\begin{table}
\caption{Periods and Mode Identification.}
\label{tab:fit}
\begin{center}
\begin{tabular}{lccccll}
\hline
      &         & \multicolumn{3}{c}{$P_{\rm calc}$} &     &   \\
\cline{3-5}
 f$_{\rm num}$  & $P_{\rm obs}$ & Pure C & 50:50 C/O & Pure O  & $\ell$ & n \\
\hline
 f$_{12}$ & 195.0 & 192.57 & 192.82 & 192.76 & 2   &  6    \\
 f$_{10}$ & 204.6 & 205.23 & 205.72 & 205.86 & 1   &  3    \\
 f$_{8}$  & 256.9 & 260.37 & 260.61 & 260.85 & 2   &  9    \\
 f$_{6}$  & 281.0 & 280.84 & 281.08 & 281.13 & 1   &  5    \\
 f$_{4}$  & 333.5 & 330.89 & 330.65 & 330.70 & 2   &  12   \\
 f$_{3}$  & 350.6 & 350.44 & 350.44 & 350.39 & 2,1 &  13,7 \\
 f$_{2}$  & 525.4 & 523.98 & 524.47 & 524.60 & 2   &  21   \\
 f$_{1}$  & 539.8 & 540.92 & 541.01 & 541.27 & 2   &  22   \\
\hline
\end{tabular}
\end{center}
\end{table}

Given that the analysis in the previous subsection did not uncover any
clear trends in the pulsation mode data (except for perhaps the
rotational and magnetic splitting we have proposed), one would hope
that use of a global fitting procedure could yield some definitive
results.  Adopting this mode splitting model reduces the 11 identified
normal modes to only 8 independent modes with different $n$ and/or
$\ell$ values. This is comparable to the number of modes available for
the analyses of both GD\,358 and CBS\,114 discussed above.  But, in
contrast to these other two pulsators where additional constraints are
evident ($\ell = 1$ assignments with consecutive radial index values), the
EC\,20058 pulsation spectrum offers none of these clues.  The revealed
modes are most likely a mixtures of $\ell=1$ and $\ell=2$.

Using the 8-mode data set, we applied a modified version of the global
model-fitting procedure originally described by \cite{mmk03}. This
version of the code incorporates the OPAL radiative opacities
\citep{ir96} rather than the older LAO data \citep{hue77}, which are
known to produce systematic errors in the derived temperatures
\citep{fb94}. The fitting procedure uses a parallel genetic algorithm
\citep{mc03} to minimize the root-mean-square residuals between the
observed and calculated periods ($\sigma_{\rm P}$) for models with
effective temperatures ($T_{\rm eff}$) between 20\,000 and 30\,000~K,
and stellar masses ($M_{\star}$) between 0.45 and 0.70 $M_\odot$. We
restricted the mass range more than in earlier applications to avoid a
family of models with high masses, which contain such a high density
of $\ell=2$ modes that they can match essentially any set of observed
periods. We allowed the base of the uniform He/C envelope to be
located at an outer mass fraction $\log(M_{\rm env}/M_{\star})$
between $-2.0$ and $-4.0$. The base of the pure He surface layer could
assume values of $\log(M_{\rm He}/M_{\star})$ between $-5.0$ and
$-7.0$.

Since we have almost no information about the spherical degree of the
modes, we assumed only $\ell=1$ and $\ell=2$ modes were observable and
calculated all of the periods in the range 150$-$600\,s for each
model; we then selected the closest model period for each observed
mode.  This procedure tends to bias the identification in favor of
$\ell=2$ modes since there are always more of them for a given model.
Following \cite{mmk04}, we required an $\ell=2$ mode to be closer to
the observed period by a factor ($N_{\ell=2}/N_{\ell=1}$) for
selection as a better match. In effect, we optimized the mode
identification internally for each model evaluation, while the genetic
algorithm optimized the values of the other four parameters.

To quantify the effect of our ignorance of the core composition, we 
repeated this fitting procedure using three different types of cores: pure 
C, a uniform 50:50 mixture of C/O, and pure O. The results of these three 
fits are shown in Table~\ref{tab:param}, and the corresponding model periods 
and mode identifications are shown in Table~\ref{tab:fit}. With the exception 
of the period at 350.6\,s, the mode identifications are the same for all 
three fits. The fit using a pure C core identified the 350.6\,s period as 
($\ell=1, n=7$) while the other two fits both preferred an identification 
of ($\ell=2, n=13$). The pure C model also includes this $\ell=2$ mode, 
and although it was closer to the observed period than the $\ell=1$ mode, 
it was not close enough to overcome the requirement that it be closer by 
the factor ($N_{\ell=2}/N_{\ell=1}$). Thus, the mode identification 
appears to be reasonably robust.

All three core compositions yield very good fits to the observed periods, 
with $\sigma_{\rm P}\sim2$ seconds in all cases. This is slightly better 
than the quality of the 4-parameter fits achieved for the DBV white dwarfs 
GD\,358 and CBS\,114 by \cite{met05} using the same code. The main effect of 
the different core compositions is to modify the optimal value of $T_{\rm 
eff}$, and to make slight adjustments to the locations of the two 
near-surface composition gradients. The average values of the mass and 
effective temperature are in good agreement with the spectroscopically 
determined values \citep{bea99}, the total envelope mass is within the 
range expected from stellar evolution theory \citep{dm79}, and the 
thickness of the pure He surface layer, when compared to similar measures 
for GD\,358 and CBS\,114, is consistent with the expectations of diffusion 
theory \citep{met07}.

\subsection{Review of the global fitting}

At this point it is prudent to pause and critically examine where our
analysis has led us.  From a total of 11 detected pulsation modes, we
assumed that 8 of them had independent $n$ and $\ell$ values, and then
our modelling has identified unique index values for all but one of
these modes (Table~\ref{tab:fit}).  An obvious question is how
plausible and unique is our `best' model?

One interesting result, which was foreshadowed in our previous general
discussion, is the assignment of different $\ell$ values (1 and 2) to
the dominant 281\,s (f$_6$) and 257\,s (f$_8$) equal (observed)
amplitude modes.  Taking into account the expected larger
geometrical flux cancellation from an $\ell=2$ mode, this requires a
higher physical amplitude for the 257\,s mode to compensate for this
geometrical effect.  In isolation, a more plausible
assumption might be that both modes have the same $\ell=1$ value, and
the small period spacing between them is explained by an unusually
high mass ($\sim$ 0.8\,M$_{\odot}$) for the star.  But, we do have
external constraints: model fits to EC\,20058's optical spectrum are
not consistent with such a high mass \citep{bea99,sul07}.  It is also
interesting that a high mass can be ruled out largely through
asteroseismic constraints, as discussed in the next paragraph.

As a further check on our conclusions, we undertook a relatively
exhaustive exploration of model parameter space using a grid of
pre-computed models.  Unconstrained fitting to the pulsation spectrum
using criteria similar to that discussed previously yielded two
competing best fit models: the `low-mass' ($\sim$ 0.55\,M$_{\odot}$)
ones discussed previously, and a `high-mass' family with M $\sim$ 0.8
-- 0.9\,M$_{\odot}$.  However, when the 281\,s (f$_6$) mode was
constrained to be $\ell = 1$, then the high-mass family did not remain
competitive, and hence the best model assigned $\ell = 2$ to the
257\,s mode and predicted a mass of 0.55\,M$_{\odot}$, confirming the
above.  Assigning $\ell = 1$ to the 281\,s mode is not totally
arbitrary and is consistent with the rotational and magnetic mode
splitting arguments developed in the previous section.

Thus, two independent constraints added to the asteroseismic fitting
resulted in the same conclusion.  We conclude our modelling discussion
at this point, and add the comment that a more exhaustive modelling
effort is currently underway.

\section{Summary and future work}

The work presented here has significantly extended our knowledge of
the helium atmosphere pulsating white dwarf, EC\,20058$-$5234 beyond
that revealed by the discovery observations of \citet{koe95}.
Analysis of nearly 135\,h of photometric time series data obtained by
four southern telescopes during the 8-day 1997 WET run {\sc xcov}15,
along with 42\,h of single site data obtained during a week of
observing at Mt John Observatory in June 2004 has more than doubled
the number of detected frequencies from the 8 to 18 (Table~\ref{tab:freq}).

Following a detailed simulation procedure that we undertook, we
are confident at the 1 in 1000 false alarm probability level that the
power in the detected modes (the lowest 0.6 mma frequencies, in
particular) is not simply the result of random noise fluctuations.

Although we were able to identify signal power in our WET data at 18
different frequencies that are above our adopted 0.54 mma significance
level (and one possibility just below: `$a$'), only 11 of these
could reasonably be viewed as pulsation modes.  The 7 other periods
are readily characterised as combination frequencies resulting from
the nonlinear processes that translate mechanical movement of the
stellar material into observable flux changes.

Our initial investigation of the detected pulsation mode structure
revealed no clear systematic trends, except for perhaps a rotational
splitting value of 70 (or 75) $\mu$Hz for two of the modes.
Interestingly, the modes suspected of showing rotational splitting
were independently identified as the only $\ell=1$ modes from our
objective fitting method, which could explain why they are the only
modes that show this behavior.  We (tentatively) interpret this as
evidence that the white dwarf is rotating with a period of about 2
hours.

A global fitting analysis of the period structure, following
procedures similar to those described in \citet{met00}, resulted in 3
models with different core compositions that all yielded very good
fits to the 8 assumed independent observed periods (see
Tables~\ref{tab:param} and \ref{tab:fit}).  All models had similar
effective temperatures ($\sim$ 28\,200\,K) and masses (0.55
M$_{\sun}$) that were in good agreement with the spectroscopically
determined values \citep{bea99,sul07}.  This analysis provided no
insights on the core composition, as the fits were largely insensitive
to the 3 different chosen C/O ratios.  As expected from the earlier
discussions, the global fitting assigned different mode $\ell$ values
to the two dominant pulsation frequencies: f$_{6}$, $\ell=1$ and
f$_{8}$, $\ell=2$.

Although the number of known DBV objects is relatively small (17), and
is totally dwarfed by the current known number of cooler hydrogen
atmosphere (DAV) pulsators (now exceeding 140), it is still tempting
to draw some conclusions about the nature and extent of the DB
instability strip along the lines discussed by \citet{muk04b} (see
also \citealt{muk06}).  Even using just the two stars GD\,358 and
EC\,20058, we can infer that the DBVs mimick the characteristics
exhibited by the DAVs.  GD\,358 has a relatively low effective
temperature (24\,900\,K, \citeauthor{bea99}) and exhibits both large
amplitude pulsations and unstable behaviour (modes come and go), while
EC\,20058 has a significantly higher effective temperature
(28\,400\,K) and displays low amplitude very stable behaviour.
Similar behaviour is found in the ensemble of DAVs, and some authors
have divided them into hot (hDAV) and cold (cDAV) pulsators
\citep[e.g.][]{muk04b}.  The hDAVs feature only a few detectable
pulsation modes that have low amplitudes and are relatively stable,
while the cDAVs exhibit many variable modes.  This is surely telling
us something about the onset and cessation of pulsation as the white
dwarf cools through the temperature dependent instability strip.

Certainly compared with GD\,358, EC\,20058 is a stable pulsator in
terms of both mode amplitude and period.  The two lower panels in
Fig.~\ref{fig:f6+f8}, in particular, demonstrate that the mode
amplitude stability might be subject to some qualifaction, as the mode
f$_{7}$ has an amplitude of 1.2 mma in the WET data, but there is no
evidence of it in the MJ04 data.  An alternative possibility is that
beating between unresolved modes explains the low-level amplitude
variability.  However, the dominant modes are always present in the
photometry with about the same amplitudes.

Both data sets presented here indicate that EC\,20058 has \emph{very}
stable pulsation periods.  This conclusion is also supported by
other observational work \citep{sul00b,sul03,sul05}.  This means that
we can attempt to measure evolutionary cooling effects over human
timescales.  The key point is that the DBV white dwarf model periods
increase with decreasing temperature \citep[e.g.][]{bra93}, so careful
monitoring of a stable mode period or periods over a multi-year interval
should in principle lead to a detectable period change and therefore a
direct measure of the stellar evolutionary timescale.

As one might expect, the models for both the cooler DAV white dwarfs
and the much hotter DOV pulsating pre-white dwarfs also predict period
increases with decreasing temperature.  More than 10 years of
time-series photometry (including several WET runs) on the DOV object
PG\,1159$-$035 produced a measured period increase \citep{cos99},
largely because the predicted rate is several orders of magnitude
higher than for the DBVs.  Note, that as a direct testimony on the
difficulties of these measurements, a `false alarm' result was
published at an earlier date \citep{win85,win91}.  Kepler and
collaborators \citep{kep05a} have been monitoring the highly stable
DAV G\,117$-$B15A over three decades and have finally detected a
definite period change.  As recently pointed out by \citet{win04},
with sufficient quality data one might expect to measure a period
change for a DBV over a timescale as short as $\sim$ 5 years.  This is
the aim of an ongoing observational campaign featuring EC\,20058
\citep{sul05}.

The cooling mechanism for both the very hot pre-white dwarfs and the
hot white dwarfs is dominated by neutrino emission from the
high-temperature degenerate cores \citep[e.g.][]{win83,obr00b,win04}.
Models show that this core neutrino flux exceeds the surface photon
flux down to an effective temperature of about 25\,000\,K, depending
on the mass of the star.

In fact, \citet{win04} demonstrate that a DB model with a typical mass
of 0.6\,M$_{\odot}$ with an effective temperature of 28\,400\,K (the
EC\,20058 estimated value), has a neutrino cooling flux approximately
four times that of the photon flux.  These facts suggest a very
interesting experiment: measure a period change and compare this rate
with two values obtained from models that include and exclude the
neutrino mechanism.  Then, provided we are confident that the star's
mass and effective temperature are in the right region, we have the
exciting possibility that we are using a pulsating star to directly
test a key low-energy prediction of electroweak theory.

This is the current promise for EC\,20058 as at these effective
temperatures the evolutionary cooling occurs at approximately constant
radius, in contrast to the much hotter pre-white dwarfs
\citep*{win83}.  In the latter case, both continuing gravitational
contraction and evolutionary cooling have an impact on the pulsation
periods so any comparisons with theoretical predictions are less certain.

Given the potential importance of this star, we have recently
undertaken both time-series photometry and spectroscopy using one of
the Magellan 6.5\,m telescopes \citep{sul07}, in addition to seasonal
monitoring using Mt John Observatory \citep{sul03}.

\section*{Acknowledgments}
We thank the various Time Assignment Committees for the award of
telescope time, and in particular, DJS would like to thank the
Department of Physics and Astronomy at the University of Canterbury
for the generous allocation of Mt John observing time for this
project. He also thanks Grant Kennedy for help in reducing the MJ04 data
set and the VUW Faculty of Science for financial support.  We also
thank Scot Kleinman for a very thorough reading of initial drafts of
the paper. TSM would like to thank Mark Waddoups for his assistance
with the model fitting.  Finally we thank an anonymous referee for
careful reading of the manuscript and constructive comments that
led to an improvement in the clarity of the presentation.

\bibliographystyle{mn2e}
\bibliography{wd_bibliography}

\label{lastpage}

\end{document}